\documentclass{jfm}

\usepackage{natbib}
\usepackage{hyperref}
\usepackage{graphicx}
\usepackage{epstopdf, epsfig}
\usepackage{amsmath}
\usepackage{esint}
\usepackage{float}
\usepackage{caption}
\usepackage{subcaption}
\usepackage[utf8]{inputenc}
\usepackage[export]{adjustbox}
\usepackage{wrapfig}
\usepackage{hyperref}
\usepackage{float}
\usepackage{xcolor}

\shorttitle{Viscous flow-fields in hyperelastic Chambers}

\title{Viscous flows in  hyperelastic chambers}

\author{Eran Ben-Haim,
   Dotan Ilssar, Yizhar Or
 \and Amir D. Gat  \corresp{\email{amirgat@technion.ac.il}}}

\affiliation{Faculty of Mechanical Engineering, Technion - Israel Institute of Technology, Haifa 3200003, Israel}

\begin{document}
\maketitle
\begin{abstract}
Viscous flows in hyperelastic chambers are relevant to many biological phenomena such as inhalation into the lung's acinar region and medical applications such as the inflation of a small chamber in minimally invasive procedures. In this work, we analytically study the viscous flow and elastic deformation created due to inflation of such spherical chambers from one or two inlets. Our investigation considers the shell's constitutive hyperelastic law coupled with the flow dynamics inside the chamber. For the case of a narrow tube filling a larger chamber, the pressure within the chamber involves a large spatially uniform part, and a small order correction. We derive a closed-form expression for the inflation dynamics, accounting for the effect of elastic bi-stability. 
Interestingly, the obtained pressure distribution shows that the maximal pressure on the chamber's surface is greater than the pressure at the entrance to the chamber. The calculated series solution of the velocity and pressure fields during inflation is verified by using a fully coupled finite element scheme, resulting in excellent agreement. Our results allow estimating the chamber's viscous resistance at different pressures, thus enabling us to model the process of inflation and deflation. 
\end{abstract}


\section{Introduction}\label{sec:Introduction}
The inflation of elastic balloons has been extensively investigated in the past, mainly because the corresponding dynamics depend on both the flow and the balloon's material elasticity model. The inflation of a toy balloon or a spherical membrane was studied thoroughly by \cite{Beatty}. In his work, \cite{Beatty} has presented an analysis of an incompressible, isotropic hyperelastic spherical pressurized membrane. According to his work, and  similar results by \cite{Treloar}, the Mooney-Rivlin elasticity model successfully captures most of the overall physical effect. The majority of the research done so far considered hydrostatic uniform pressure distribution within the chamber and the determination of pressure as a constant parameter that uniformly affects the elastic walls \citep{Needleman,Treloar,Beatty,Vandermarliere,Hines,Mangan}. 

Balloons with controlled inflation are used in medical applications such as pleural pressure assessments \citep{Milic-Emili}, and enteroscopy \citep{Yamamoto}. A recent study by \cite{Manfredi} shows a promising biomedical application of a soft robot for a colonoscopy, which utilizes a double-balloon system for achieving inchworm-like crawling while bracing against the colonic walls. \cite{ Haber} investigated alternating shear flow over a self-similar, rhythmically expanding hemispherical depression. Quasi-steady creeping flow in models of small airway units of the lung was investigated by \cite{Davidson}. \cite{Ilssar} studied the inflation and deflation dynamics of a liquid-filled hyperelastic balloon, focusing on inviscid laminar flow. In those systems, the characteristic time it takes for the pressure to reach a constant uniform value in a chamber is assumed to be much shorter than the time it takes for the fluid to pass through the tubes (based on the viscous resistance). However, to assess the fluid and elastic shell's dynamics, a complete mathematical model describing the system's fluid-structure interaction at Low-Reynolds numbers is needed. The study of the fluid-structure interaction dynamics of low-Reynolds-number incompressible liquid flows and elastic structures may help introduce a new level of control in fluid-structure based autonomous systems due to the presence of viscous force \citep{Elbaz14,Elbaz16}.

In the Soft-Robotics field, recent studies show the propulsion of elastic structures embedded with internal cavities while controlling pressures or flow rates at the network's inlets \citep{BenHaim,Salem,Gamus,Siefert,Fei14,Fei16,Overvelde,Gorissen}. In the case of fluidic actuation, several works study variations of the well-known `two-balloon system', whereas others study networks of multiple connected chambers \citep{BenHaim,Dreyer,Treloar,Glozman}. As shown by these studies, for some given values of the pressure, multiple solutions for the volume are possible. Since the hyperelastic spherical membranes are multi-stable systems, it allows us to selectively inflate each balloon to one of its stable states by varying the input according to a particular carefully synthesized profile. Consequently, it can pave the way toward manufacturing soft robots that utilize minimal actuation to produce highly complex locomotion.

In this work, we examine the effect of elasticity on transient creeping flow in the bi-stable hyperelastic chambers. The chamber is assumed to be an ideal sphere. Stokes equation governs the flow field, while Mooney-Rivlin constitutive laws model the elastic chamber. The fluidic pressure within the balloon is not uniform and cannot be directly determined from the known Mooney-Rivlin relation. The fluidic pressure distribution in the chamber is estimated by balancing the fluidic pressure with the total force on the elastic membrane. Since this force is obtained by integrating the pressure distribution (which depends on the angle $\theta$), it receives a different value than the value obtained in the hydrostatic models. 

This work's structure is as follows: in § \ref{sec:Problem_Formulation}, the geometry, relevant parameters and physical assumptions are defined. In § \ref{ElsticSection}, the hyperelastic Mooney-Rivlin constitutive law is presented. The strain energy function is analyzed in order to present the bi-stable phenomena. Section § \ref{sec_Solution} presents closed-form solutions of the governing equations, describing the flow field within an expanding chamber. In § \ref{sec:Results}, two different physical cases are described. The case of dictated inlet mass rate coupled by the hyperelastic model is described in § \ref{Dynamic_case_II}, where numerical verification of the fully coupled model is presented. In § \ref{Dynamic_case_III}, we present the second case where the inlet pressure is dictated and the stretching of the hyperelastic sphere is governed by the flow dynamics. Section § \ref{TwoBalloons}  examines the dynamic behavior of two interconnected bi-stable chambers. Concluding remarks are presented in § \ref{Concludings}.

\section{Problem Formulation}\label{sec:Problem_Formulation}
\begin{figure}
  \centering
    \includegraphics[width=1.0\textwidth]{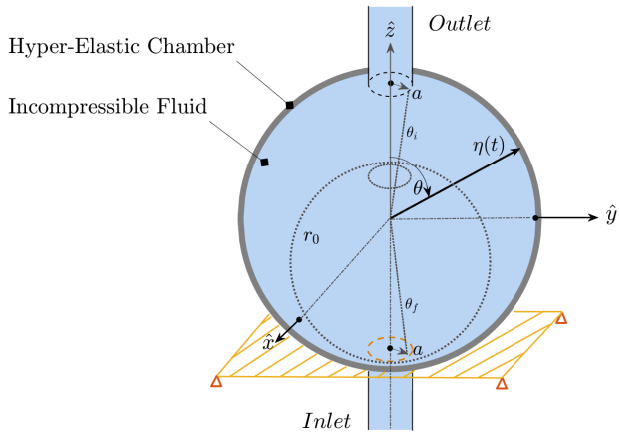}
    \caption{Illustration of the system under investigation, consisting of a hyper-elastic chamber, as well as a fixed inlet tube, and a massless outlet tube, both having identical radius and length. The bottom of the chamber is held fixed during the inflation, while its center is allowed to move. }
    \label{fig:Illustration}
\end{figure}

In this section, we present the problem definition, along with the physical parameters relevant to the analysis and the small non-dimensional parameters. The examined liquid-filled chamber is illustrated in Figure \ref{fig:Illustration}. Here, a spherical geometry is assumed (the validity of this assumption will be verified by numerical simulation in \S \ref{Dynamic_case_II}). A spherical hyperelastic chamber with a stress-free radius of $r_{0}$, is connected to two rigid tubes with radius $a$ and length $\ell$. For simplicity, we assume identical tubes in the inlet and the outlet. Here, the flow-field inside the chamber and tubes is considered incompressible, Newtonian, and with negligible inertial effects. The fluid's axial velocity inside the tube is $u_{z}$, and the volumetric flux rate is denoted $q(t)$ (where $q_{in}(t)$ refers to the flow entering the body from the inlet tube and $q_{out}(t)$ refers to the flow moving from the body through the outlet tube). The relevant variables and parameters are the time $t$, the axial coordinate and symmetry axis $z$, and the radial coordinate $s$ of the cylindrical system used to describe the tubes. Axisymmetry allows us to eliminate the azimuthal angle of the cylindrical system. Furthermore, the pressure and flow velocity fields of the entrapped fluid are $p(t,s,z)$ and $\textbf{v}(t,s,z)$, while its constant density and dynamic viscosity are denoted $\rho$ and $\mu$. The chamber's dynamics are approximated by a single degree of freedom, represented here by the chamber's instantaneous radius, denoted $\eta(t)$.  For spherical geometry, a coordinate system is chosen so that one of the coordinates remains constant on the boundary. Here, $\{r,\theta,\phi\}$ are the coordinates of a moving spherical system, located at the center of the chamber, where $\theta$ is the polar angle, measured from the axis of symmetry to the radial coordinate $r$, and $\phi$ is the azimuthal angle, revolving around the axis of symmetry, $z$. The Cauchy-stress tensor of the flow is denoted as $\mathbf{\sigma}_\mathbf{f}$. The stress-free shell's thickness is $w_{0}$, which is considered to be much smaller than the stress-free chamber's radius, namely $w_{0}\ll r_{0}$.

The following analysis utilizes three small parameters, including the ratio between the radius of the tubes and the radius of the stress-free chamber ($\varepsilon_a$, denoted hereafter by \textit{tube-chamber radii ratio}),
\begin{equation}
    \epsilon_{a} = \cfrac{a}{r_{0}}\ll 1,
\end{equation}
the slenderness of the tubes ($\varepsilon_t$, denoted hereafter by \textit{tube slenderness}),
\begin{equation}
    \epsilon_{t} = \cfrac{a}{\ell}\ll 1,
\end{equation}
and the last small parameter in the analysis is taken as the ratio between the viscous stresses and the overall pressure in the chamber ($\varepsilon$, denoted hereafter by \textit{chamber viscous resistance parameter}), defined by
\begin{equation}
    \varepsilon = \cfrac{\mu v^{*}}{r_0 p^{*}}\ll 1,
    \label{epsi}
\end{equation}
where $v^{*}$ is the characteristic flow velocity in the chamber, and $p^{*}$ is the characteristic pressure of the system.
For the following analysis, we shall normalize the physical variables by considering the characteristic values of the problem as follows:
\begin{equation}
    \textbf{V}=\cfrac{\boldsymbol{v}}{v^*},\quad \hat{\boldsymbol{\nabla}}=r_{0}\boldsymbol{\nabla},\quad \hat{\boldsymbol{\sigma_{f}}}=\cfrac{\boldsymbol{\sigma_{f}}}{p^{*}},\quad R=\cfrac{r}{r_{0}},\quad \lambda=\frac{\eta}{r_{0}},\quad P=\cfrac{p}{p^{*}},\quad T=\cfrac{t}{r_{0}/v^{*}},
    \label{normalized} 
\end{equation}
where $\lambda(T)$ denotes the stretch of the chamber, and $R$ is the normalized radial coordinate.

\section{Constitutive model for a hyperelastic membrane}
\label{ElsticSection}
This section presents the constitutive law that governs the spherical shell dynamics. We consider a thin-shelled, spherical chamber made of incompressible hyperelastic isotropic material. Finite elasticity theory dictates a known form of the elastic strain energy density $\psi(\lambda)$, which depends only on the relative stretch $\lambda(T)$. Moreover, the elastic strain energy density satisfies $\psi(1)=0$. Different types of hyperelastic models differ in the type of material and the elastic strains experienced without failing. The most common models are \textit{neo-Hookean, Mooney-Rivlin, Ogden, Gent} and \textit{Biological tissue} \citep{Ogden}.

The material is assumed to be incompressible, which leads to the relation between the pressurized and the stress-free states, given by $r^{2}w\approx r_{0}^{2}w_{0}$. Thanks to this relation, the chamber's instantaneous thickness is eliminated. To capture the chamber's bi-stability, we use the two-parameter Mooney-Rivlin model.

Under the above assumptions, and considering incompressibility, the normalized solid's Mooney-Rivlin strain energy function is given by \citep{Ogden,Beatty},
\begin{equation}
\label{MRpotential}
    \hat{\psi}(\lambda)=2\lambda^{2}+\cfrac{1}{\lambda^4}-3+\alpha\bigg(\lambda^4+\cfrac{2}{\lambda^{2}}-3\bigg)
\end{equation}
where $\alpha=s_{2}/s_{1}$ is the ratio between two empirically determined constants, commonly denoted as the Mooney–Rivlin parameters.

In this study, the Mooney–Rivlin parameters are chosen as $s_{1}\approx1.5$ MPa and $s_{2}\approx0.15$ MPa \citep{Beatty,Treloar}. The normalized equation (\ref{MRpotential}) was obtained by using the normalization $\psi^{*}=s_{1}$ for the strain energy density function, and the magnitude of the parameter $\alpha$ is $O(10^{-1})$.

We first study the chamber's static behavior, where the pressure (without flow) is dictated. In this case, both the stretch and the pressure are constant, denoted here by $\lambda_{SS}$ and $P_{SS}$, respectively. 
The behavior mentioned above can be demonstrated by the overall effective potential energy of the system,
\begin{equation}
\label{EffectivePotential}
    \mathcal{U}(\lambda_{SS};P_{SS})=\int_{\lambda_{SS}}{\bigg(\cfrac{d\hat{\psi}}{d\xi}-\xi^{2}P_{SS}}\bigg)\mathrm{d}\xi=\hat{\psi}(\lambda_{SS})-\frac{1}{3}\lambda_{SS}^{3}P_{SS}.
\end{equation}

\begin{figure}
    \centering
    \includegraphics[width=1\textwidth]{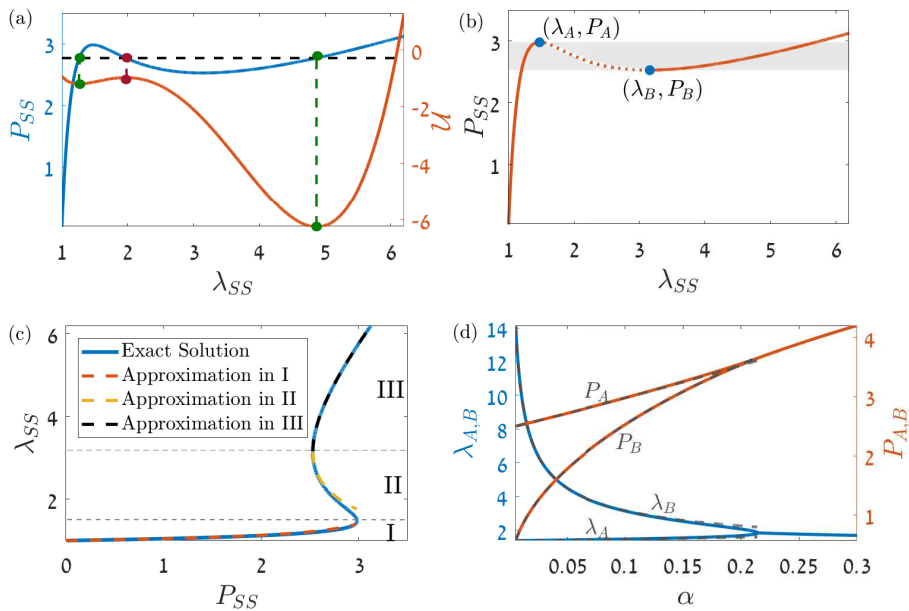}
    \caption{(a) - The solid blue curve is a characteristic stretch–pressure curve of a single elastic chamber with $\alpha=0.1$. The solid orange curve is the effective potential energy function (\ref{EffectivePotential}), corresponding to the constant pressure $P_{SS}=2.75$, illustrated by the dashed black line; Green and red dots are the stable and unstable equilibrium radii. (b) - Characteristic normalized stretch–pressure curve of a single elastic chamber according to (\ref{Pss}) with $\alpha=0.1$. The extrema are marked, and the bi-stable region is marked in grey. Solid curves are stable branches and dashed
    curves are unstable ones. (c) - The solid blue line is the exact solution of (\ref{Pss}) calculated numerically. The dashed curves are the approximated solutions obtained in (\ref{delta1Appr})-(\ref{delta2Appr}). (d) - The evolution of the extremum values of pressure $P_{A},P_{B}$ and stretch $\lambda_{A},\lambda_{B}$ as a function of the small parameter $\alpha$. The solid curves are the exact values, and the dashed curves are the asymptotic approximations given in appendix (\ref{PAapprox}),(\ref{PBappr}).}
    \label{fig:staticCase}
\end{figure}

Based on (\ref{EffectivePotential}), Figure \ref{fig:staticCase}(a) shows a curve of the potential energy function where the constant pressure is $P_{SS}=2.75$. The Mooney-Rivlin relation (\ref{MRpotential}), along with the steady version of the leading order energy balance (\ref{EffectivePotential}), formulated as $\partial\mathcal{U}/\partial\lambda_{SS}=0$ at constant pressure $P_{SS}$, yields a relation between stretch, $\lambda_{SS}$, and pressure, $P_{SS},$ in equilibrium condition,
\quad
\begin{equation}
\label{Pss}
    P_{SS}=\bigg(\cfrac{1}{\lambda^{2}}\frac{d\hat{\psi}}{d\lambda}\bigg)\bigg|_{\lambda_{SS}}=4\bigg[\frac{1}{\lambda_{SS}}-\frac{1}{\lambda^{7}_{SS}}+\alpha\bigg(\lambda_{SS}-\frac{1}{\lambda^{5}_{SS}}\bigg)\bigg]; \qquad 0<\alpha\ll 1.
\end{equation}

This well-known  relation was extensively leveraged to describe the quasi-static inflation of spherical balloons \citep{Beatty,Treloar,BenHaim} for spatially uniform pressures. As seen from Figure \ref{fig:staticCase}(a), showing the relation in (\ref{Pss}) with $\alpha=0.1$, the uniform pressure of the chamber is not monotonic with respect to the radius. Therefore, the inverse relation, describing the chamber's radius as a function of the pressure, cannot be directly extracted.  The curve $P_{SS}(\lambda_{SS})$ in Figure \ref{fig:staticCase}(b) has two bifurcation points, described by a local maximum point at $(\lambda_{A},P_{A})$, and a local minimum point at $(\lambda_{B},P_{B})$. This figure shows a bifurcation, which occurs when the pressure enters or exits the range between the local extrema, $P_{B}<P_{SS}<P_{A}$, illustrated in grey. Asymptotic approximations for the bifurcation points of the equilibrium curve, $P_{A},P_{B},\lambda_{A}$ and $\lambda_{B}$, appears in Appendix \ref{appA}. The  evolution  of  those  extrema as a function of the small parameter $\alpha$ is presented in Figure \ref{fig:staticCase}(d). Asymptotic approximations for the solution of the equilibrium equation (\ref{Pss}), appear in Appendix \ref{appA}. Those approximations are plotted in Figure \ref{fig:staticCase}(c) with dash-lines on the solid exact solution curves, represented by the inverse relation $\lambda_{SS}(P_{SS})$.

Analyzing the strain energy function $\mathcal{U}(\lambda_{SS};P_{ss})$ in Equation (\ref{EffectivePotential}), using the second derivative with respect to $\lambda_{SS}$, it can be proven that the right and left branches of $P_{SS}(\lambda_{SS})$ where $1<\lambda<\lambda_{A}$ or $\lambda_{B}<\lambda$ are stable equilibria and satisfy, 
\begin{equation}
\label{second_derivative}
    \cfrac{\partial^2\mathcal{U}}{\partial\lambda_{SS}^{2}}\bigg|_{P_{SS}}=\cfrac{dP_{SS}}{d\lambda_{SS}}>0.
\end{equation}
Conversely, the intermediate branch $\lambda_{A}<\lambda_{SS}<\lambda_{B}$ is an unstable region satisfying $\partial^2\mathcal{U}/\partial\lambda_{SS}^{2}<0$. This is precisely the \textit{bi-stability} phenomenon.

\section{Series solution of the flow-field within an expanding chamber}
\label{sec_Solution}
In this section, the governing equations of the flow within the chamber will be formulated, as well the problem's boundary conditions. An analytical series solution will then be presented, describing the velocity field and the flow's pressure distribution inside the spherical chamber. 
\subsection{\textbf{Formulation and analysis of the governing equations}}
Under the assumptions discussed above, the momentum and continuity equations governing the fluid's behavior expressed in the moving spherical frame: 
\begin{subeqnarray}
   \rho\bigg(\cfrac{\partial{\boldsymbol{v}}}{\partial{t}}+\boldsymbol{v}\bcdot\bnabla\boldsymbol{v}+\hat{\boldsymbol{z}}\cfrac{d^2}{dt^2}\sqrt{\eta^{2}-a^{2}}\bigg)& = &-\bnabla p+\mu\bnabla^{2}\boldsymbol{v}-\rho g\hat{\boldsymbol{z}} ,\\[3pt]
\bnabla\bcdot\boldsymbol{v} & = & 0 ,
\label{momentum_tube}
\end{subeqnarray}
where the third term in the left expression in the momentum equation (\ref{momentum_tube}a) describes the acceleration of the moving spherical frame, centered on the chamber's moving center, relative to a stationary frame.

Assuming the flow in tubes is fully developed and axisymmetric, the volumetric flux is given by,
\begin{equation}
    q(t)=-\cfrac{\pi a^4}{8\mu} \frac{\partial p_{t}}{\partial{z}},
    \label{Hagen-Poiseuille}
\end{equation}
where $\partial p_{t}/\partial z$ is the pressure gradient along the tube. Since the tube slenderness $\epsilon_{t}\ll 1$ we shall assume a constant pressure gradient. Normalization of (\ref{Hagen-Poiseuille}) yields the characteristic flow rate as $q^{*}=\pi a^2 u_z^{*}$ where $u_{z}^{*}=a^{2} p^{*}/\mu\ell$ is the characteristic axial component of the fluid velocity in the tube. An integral flow balance yields the relation between the characteristic velocity in the tube and the characteristic velocity of the flow within the chamber, as $v^{*}=\epsilon_{a}^{2} u_{z}^{*}$. Substituting the characteristic values into the chamber viscous resistance parameter (\ref{epsi}), relates it to the other small parameters, as follows
\begin{equation}
    \varepsilon = \frac{a^{4}}{r_{0}^{3} \ell}=\epsilon^{3}_{a}\epsilon_{t}\ll 1.
    \label{small}
\end{equation}

From relation (\ref{small}) it is clear that $\varepsilon$ is dependent merely on the geometry of the system, providing a simple relation between the hydrostatic and deviatoric stresses. Hence, an appropriate geometry can be defined in order to design an efficient and controllable system.
We consider negligible gravity, i.e., $\rho gr_{0}/p^*\ll1$ (where $g$ is the gravitational acceleration), and define a Reynolds number in chamber as $\Rey=\rho v^{*}r_{0}/\mu\ll 1$. Since the Reynolds number is small, the flow's inertia may be neglected. Therefore, by utilizing the non-dimensional quantities specified in (\ref{normalized}) the fluid's motion (\ref{momentum_tube}) is governed by Stokes equations for creeping flow with an implicit time variable,
\begin{equation}
    \hat{\bnabla}\bcdot\boldsymbol{V}=0, \quad \hat{\bnabla} P=\varepsilon \hat{\bnabla}^{2}\boldsymbol{V}+O(\varepsilon \Rey).
    \label{stokes} 
\end{equation}

The validity of these equations is weakened at the vicinity of the connections to the tubes since in those regions, the characteristic velocity is approximately $u_{z}^{*}$ rather than $v^{*}$.
At a distance $O(L)$ from the tube (where $a<L<r_{0}$), the velocity scale is $q^*/L^2$ and hence the Reynolds number is $\rho q^*/\mu L<\rho q^*/\mu a$. Thus, a sufficient condition for global neglect of inertia is,
\begin{equation}
    q^{*}\ll \cfrac{\pi\mu a}{\rho}.
\end{equation}

From the non-dimensional relation (\ref{stokes}), in the sphere, the pressure is spatially uniform at leading order and the viscous flow will generate small spatially varying corrections. Moreover, in order to get a better understanding of the chamber resistance small parameters' physical meaning, we may use the dynamical stress tensor in the fluid domain defined by the constitutive relation $\boldsymbol{\sigma_{f}}=-p\boldsymbol{I}+\mu\big[\boldsymbol{\nabla v}+(\boldsymbol{\nabla v})^{T}\big]$ where $\boldsymbol{I}$ is the $3\times 3$ unit matrix. In the most general constitutive equation, $\boldsymbol{\sigma}_{f}$ consisting of the linear and instantaneous dependence of the deviatoric stress, plus the hydrostatic stress, $-p\boldsymbol{I}$, stemming from the static pressure. Normalization of the total stress tensor yields,
\begin{equation}
    \boldsymbol{\hat{\sigma}_{f}}=-P\boldsymbol{I}+\varepsilon\big[\boldsymbol{\hat{\nabla}V}+\big(\boldsymbol{\hat{\nabla} V}\big)^{T}\big]
    \label{stress}.
\end{equation}

As one can notice from equation (\ref{stress}) the velocity field is not included at the leading-order. Mainly, the leading-order of the problem is a case of fully developed uniform pressure without any velocities. Suppose the flow is dictated by controlled pressure or flux at the inlet, the velocity is generated, and additional small deviatoric stress is created, which quasi-statically leads the system to another hydrostatic state.

The axisymmetric Stokes equations (\ref{stokes}) can be solved in spherical polar coordinates using a series expansion \citep{Happel}. Consider the non-dimensional Stokes stream function  $\Psi(R,\theta,T)$. The flow velocity components $V_{R}$ and $V_{\theta}$ are related to the Stokes stream function $\Psi(R,\theta,T)$ through
\begin{equation}
\label{PsiDefinition}
    V_{R}=\cfrac{1}{R^{2}\sin\theta} \cfrac{\partial\Psi}{\partial\theta};\quad V_{\theta}=-\cfrac{1}{R\sin\theta} \cfrac{\partial\Psi}{\partial R},
\end{equation}
where $V_{R}$ and $V_{\theta}$ are the radial and tangential velocity components, respectively. By applying the curl operator to the momentum equation (\ref{stokes}) and using several simple algebraic manipulations, the Stokes equation can be reduced to a fourth-order bi-harmonic equation obtained in terms of the Stokes stream function as follows:
\begin{subeqnarray}
    E^{2}\big(E^{2}\Psi\big) = 0    ,\qquad\qquad\\[3pt]
    \bnabla P = -\varepsilon\cfrac{\hat{\boldsymbol{\phi}}}{R\sin\theta}\times\bnabla\big(E^{2}\Psi\big),
    \label{bi-harmonic} 
\end{subeqnarray}
where in spherical coordinates,
\begin{equation}
    E^{2}=\cfrac{\partial^{2}}{\partial{R^{2}}}+\cfrac{\sin\theta}{R^{2}} \cfrac{\partial}{\partial{\theta}} \bigg(\frac{1}{\sin\theta} \cfrac{\partial}{\partial{\theta}}\bigg).
    \label{bi-harmonicOperator} 
\end{equation}

Equation (\ref{bi-harmonic}a) is solved by separation of variables, and equation (\ref{bi-harmonic}b) is solved by integration with respect to the radial and tangential directions. However, for brevity we will not present the full calculation of the solution here. A solution for the stream function in spherical coordinates is of the form,
\begin{equation}
    \Psi(R,\theta;T)=\mathcal{A}_{0}(T)+\sum_{n=2}^{\infty}\big[\mathcal{A}_{n}(T)R^{n}+\mathcal{C}_{n}(T)R^{n+2}\big]\mathcal{J}_{n}(\cos\theta) ,
    \label{GeneralPsi} 
\end{equation}
where $\mathcal{A}_{n}(T)$ and $\mathcal{C}_{n}(T)$ are unknown functions, determined by the boundary conditions, and $\mathcal{J}_{n}(\xi)$ are the \textit{Gegenbauer} functions of the first kind of order $n$ (and degree $-1/2$). \cite{Happel} have exhaustively investigated the properties of these Gegenbauer functions in connection with the hydrodynamic application. For our present purposes, their properties can be deduced most readily from their relation with the corresponding Legendre functions of the first kind $\mathcal{P}_{n}(\xi)$ as,
\begin{equation}
    \mathcal{J}_{n}(\xi)=\cfrac{\mathcal{P}_{n-2}(\xi)-\mathcal{P}_{n}(\xi)}{2n-1}=-\frac{1}{(n-1)!}\bigg(\frac{d}{d\xi}\bigg)^{n-2}\bigg(\frac{\xi^2-1}{2}\bigg)^{n-1};\qquad n\geqslant 2.
    \label{Legendre} 
\end{equation}

In the degenerate cases $n=0,1$ we define $\mathcal{J}_{0}(\xi)=1$ and $\mathcal{J}_{1}(\xi)=-\xi$,  respectively. Using the definition of the Stokes stream function (\ref{PsiDefinition}), we describe the series solution of the velocity field and pressure distribution as,
\begin{subeqnarray}
   V_{R}(R,\theta;T) & = & -\sum_{n=2}^{\infty}\big[\mathcal{A}_{n}(T)R^{n-2}+\mathcal{C}_{n}(T)R^{n}\big]\mathcal{P}_{n-1}(\cos\theta)   ,\\[3pt]
    V_{\theta}(R,\theta;T) & = & \sum_{n=2}^{\infty}\big[n\mathcal{A}_{n}(T)R^{n-2}+(n+2)\mathcal{C}_{n}(T)R^{n}\big]\cfrac{\mathcal{J}_{n}(\cos\theta)}{\sin\theta}   ,\\[3pt]
    P(R,\theta;T) & = & P_{0}(T)-\varepsilon\sum_{n=2}^{\infty}\bigg[\frac{2(2n+1)}{n-1}R^{n-1}\mathcal{C}_{n}(T)\bigg]\mathcal{P}_{n-1}(\cos\theta),
\label{GeneralSolution}
\end{subeqnarray}
where $P_{0}(T)$ should be determined through the physical boundary conditions defined by pressure at the chamber's inlet and outlet. The unknown functions, $\mathcal{A}_{n}(T)$ and $\mathcal{C}_{n}(T)$, should be determined by requiring the kinematic boundary conditions of the flow to be satisfied. Note that the series solutions (\ref{GeneralSolution}) are an exact solution of (\ref{bi-harmonic}).
\subsection{\textbf{Formulation and discussion of the boundary conditions}}
Next, we are interested in finding the unknown functions, $\mathcal{A}_{n}(T)$ and $\mathcal{C}_{n}(T)$ by imposing two boundary conditions. The first boundary condition is obtained from the assumption that there is no penetration into the chamber's boundaries, and the second boundary condition is the no-slip condition. The Eulerian description of a material point located on the chamber's wall is given by $\textbf{r}(\theta,t)=\eta(t)\boldsymbol{\hat{r}}$, where $\boldsymbol{\hat{r}}=(\sin\theta\cos\phi,\sin\theta\sin\phi,\cos\theta)$ is the radial direction unit vector; thus, the material point's velocity is $\boldsymbol{\dot{\hat{r}}}=\dot{\eta}\boldsymbol{\hat{r}}+\eta\dot{\theta}\boldsymbol{\hat{\theta}}$, where $\boldsymbol{\hat{\theta}}=(\cos\theta\cos\phi,\cos\theta\sin\phi,-\sin\theta)$ is the polar direction unit vector. We recall that the fluid's behavior is described in the moving spherical frame; therefore, the first term is a radial component, while the polar component is created due to the constraint of the rigid inlet/outlet tube. Assuming that the deformation is spherical, the following kinematic constraint must be satisfied,
\begin{equation}
\label{keshet}
  \cfrac{\theta-\kappa\theta_i(t)}{\theta_f(t)-\theta}=\cfrac{\theta(0)-\kappa\theta_i(0)}{\theta_f(0)-\theta(0)},
\end{equation}
where $\kappa$ indicates whether there are both inlet and outlet tubes (when $\kappa=1$), or only an inlet tube (when $\kappa=0$). Moreover, $\theta_{i}(t)$ and $\theta_{f}(t)$ are angles corresponding to the connections between the tubes and the chamber (see Figure \ref{fig:Illustration}). By simple geometric considerations, we get,
\begin{equation}
\theta_{i}(t)=\sin^{-1}\bigg(\cfrac{a}{\eta(t)}\bigg), \quad \theta_{f}(t)=\pi-\sin^{-1}\bigg(\frac{a}{\eta(t)}\bigg).
\label{thetaij} 
\end{equation}

Considering the derivative of equation (\ref{keshet}) with respect to time will lead to the relation between $\dot{\theta}$ and $\theta,\theta_i,\theta_f,\dot{\theta}_i$ and $\dot{\theta}_f$. Therefore, the material point's velocity can be rewritten as
\begin{equation}
\boldsymbol{\dot{\hat{r}}}(\theta,t)=\boldsymbol{\hat{r}}\cfrac{d\eta}{dt}+\boldsymbol{\hat{\theta}}\cfrac{(\theta_{f}-\theta)\kappa\dot{\theta}_{i}+(\theta-\kappa\theta_{i})\dot{\theta}_{f}}{\theta_{f}-\kappa\theta_{i}}\eta(t).
\label{pointVelocity} 
\end{equation}

The result obtained in (\ref{pointVelocity}) represents the change in the angle of a material point relative to the initial state.
Consequently, the no-penetration and the no-slip conditions are defined in vector form as,
\begin{equation}
\label{eq:BC}
  \boldsymbol{v}\big(r=\eta(t),\theta;t\big) =  \boldsymbol{\dot{\hat{r}}}(\theta,t)+ \boldsymbol{\hat{z}}\left\{
    \begin{array}{ll}
      \kappa u_{z}^{(out)}(s_{out}=\eta\sin\theta) & , 0\leqslant\theta\leqslant\theta_{i} \\[2pt]
     0 & ,\theta_{i}<\theta<\theta_{f} \\[2pt]
      u_{z}^{(in)}(s_{in}=\eta\sin\theta) & ,\theta_{f}\leqslant\theta\leqslant\pi
  \end{array} \right.
\end{equation}

Here, we assumed a spherical surface at the connection between the tubes and the chamber. The first vector in equation (\ref{eq:BC}) is the spherical surface velocity that captures the inflation of the whole chamber, while the second component adds the fluid's velocity into or out from the tube. Note that the velocity of the chamber's center (the velocity of the moving spherical frame, centered on the chamber’s moving center,
relative to a stationary frame) should be subtracted by the fluid's velocity components that come into or out of the tube. However, the velocity of the moving spherical frame is negligible relative to the velocity of the flow in the tube. The wall's elasticity is expressed in the body's ability to increase the volume according to the material's constitutive laws (hyper-elasticity).

A simple investigation of the boundary condition's derivative shows singularities at $\theta=\theta_{i}$ and $\theta=\theta_{f}$ where the known parabolic Hagen-Poiseuille relation is used to describe the flow inside the tube. Since the pressure distribution represented by the Legendre series (\ref{GeneralSolution}) is the solution of the second-order differential equation, singularities are leading to a divergent series. In order to avoid the singularities and to get a modified velocity profile that also takes into account the end effects, we assume a modified flow profile at the chamber's inlet (or outlet) as follows:
\begin{equation}
u_{z}^{(m)}(s,t;\epsilon_\omega) = u_{z}^{p}(s,t)+u_{z}^{c}(s,t;\epsilon_\omega).
    \label{ModifiedVelocity}
\end{equation}

Here, $u_{z}^{p}(s,t)$ is the known parabolic Hagen-Poiseuille profile
\begin{equation}
u_{z}^{p}(s,t) = \cfrac{2q(t)}{\pi a^2}\bigg[1-\bigg(\frac{s}{a}\bigg)^2\bigg],
    \label{HPVelocity}
\end{equation}
and $u_{z}^{c}(s,t;\epsilon_\omega)$ is an correction profile defined by
\begin{equation}
u_{z}^{c}(s,t;\epsilon_\omega) = \big(\omega(s;\epsilon_\omega)-1\big)u_{z}^{p}(s,t)+\gamma\cdot\cfrac{q(t)}{\pi a^2}\bigg[1-4\bigg(\frac{s}{a}\bigg)^2+3\bigg(\frac{s}{a}\bigg)^4\bigg]\omega(s;\epsilon_\omega),
    \label{CorrectionVelocity}
\end{equation}
where $\gamma$ is a parameter determined by fitting the velocity profile obtained in a finite elements calculation. The function $\omega(s;\epsilon_\omega)$ is defined by $\omega(s;\epsilon_\omega)=-\tanh{(2/\epsilon_\omega)}+\tanh{\big((s/a+1)/\epsilon_\omega\big)}-\tanh{\big((s/a-1)/\epsilon_\omega\big)}$, and $0<\epsilon_\omega\ll1$ is an arbitrary small parameter (denoted hereafter by \textit{smoothing parameter}). In fact, $\omega(s;\epsilon_\omega)$ is a weighted function making the velocity profile differentiable even at $\theta_i$ and $\theta_{f}$. Moreover, it can be shown that $\omega(s;\epsilon_\omega)=1+O(e^{-1/\epsilon_\omega})$ as $\epsilon_\omega\ll 1$ and $|s/a|\lesssim 1-O(\epsilon_\omega)$. The modified flow profile has four physical properties: the first is symmetry around the tube's radial coordinate, $s$. The second is the no-slip condition represented mathematically by zero velocity on the tube boundaries, $u_{z}^{(m)}(s=a,t)=0$. The third nulls the radial velocity gradient at the boundaries, $\partial u_{z}^{(m)}/\partial s=0$  where $s=a$. By neglecting $O(e^{-1/\epsilon_\omega})$ terms, the modified flow profile's fourth property is keeping the total flow equal to the Hagen-Poiseuille model's value (regardless of the choice of $\gamma$).
Using non-dimensional parameters, the boundary conditions become
\begin{subeqnarray}
  V_{R}\big(R=\lambda\big)=\frac{d\lambda}{dT} &+& \frac{\cos\theta}{\lambda^{2}\tilde{\epsilon}^{2}} \left\{
    \begin{array}{ll}
      \kappa U_{z}^{(m)}\big|_{S=\tilde{\epsilon}^{-1}\sin\theta} & ,0\leqslant\theta\leqslant\sin^{-1}(\tilde{\epsilon}) \\[2pt]
     0 & ,\sin^{-1}(\tilde{\epsilon})<\theta<\pi-\sin^{-1}(\tilde{\epsilon}) \\[2pt]
      U_{z}^{(m)}\big|_{S=\tilde{\epsilon}^{-1}\sin\theta} & ,\pi-\sin^{-1}(\tilde{\epsilon})\leqslant\theta\leqslant\pi
  \end{array} \right.\qquad\qquad \\[10pt]
  V_{\theta}\big(R=\lambda\big)= \Gamma(\theta) &-&\frac{\sin\theta}{\lambda^{2}\tilde{\epsilon}^{2}}  \left\{
    \begin{array}{ll}
      \kappa U_{z}^{(m)}\big|_{S=\tilde{\epsilon}^{-1}\sin\theta} & ,0\leqslant\theta\leqslant\sin^{-1}(\tilde{\epsilon}) \\[2pt]
     0 & ,\sin^{-1}(\tilde{\epsilon})<\theta<\pi-\sin^{-1}(\tilde{\epsilon}) \\[2pt]
      U_{z}^{(m)}\big|_{S=\tilde{\epsilon}^{-1}\sin\theta} & ,\pi-\sin^{-1}(\tilde{\epsilon})\leqslant\theta\leqslant\pi
  \end{array} \right.\qquad
  \label{Bo}
\end{subeqnarray}
where,
\begin{equation}
\Gamma(\theta;\tilde{\epsilon})=\Theta(\theta)\cfrac{\lambda\tilde{\epsilon}}{\sqrt{1-\tilde{\epsilon}^2}}\frac{d\lambda}{dT};\qquad \Theta(\theta)=\cfrac{(1+\kappa)\theta-\kappa(\theta_i+\theta_f)}{\theta_f-\kappa\theta_i},
\end{equation}
$\boldsymbol{U_{z}^{(m)}=u_{z}^{(m)}}/u_{z}^{*}$ is the normalized modified axial velocity in tube, $S=s/a$ is the normalized cylindrical radial coordinate in tube, and $\tilde{\epsilon}(T)=\epsilon_{a}/\lambda(T)\ll 1$. 

The integral mass conservation equation is given by,
\begin{equation}
\label{integralMass}
   q_{in}-\kappa q_{out}=\cfrac{d}{dt}\bigg[\cfrac{(1+\kappa)\pi a^2}{3}\sqrt{\eta^2(t)-a^2}+\int_{\phi=0}^{2\pi}\int_{\theta=\kappa\theta_{i}(t)}^{\theta_{f}(t)}\int_{r=0}^{\eta(t)}r^2\sin\theta\mathrm{d}r\mathrm{d}\theta\mathrm{d}\phi\bigg].
\end{equation}
We neglect $O(\tilde{\epsilon}^{4})$ terms, related to the inlet and the outlet section; thus, equation (\ref{integralMass}) is normalized and simplified to 
\begin{equation}
\label{Q-dT}
    \cfrac{d\lambda}{dT}=\cfrac{1}{4\lambda^2}\big[Q_{in}(T)-\kappa Q_{out}(T)\big].
\end{equation}

The time-dependent unknown functions in equation (\ref{GeneralSolution}), $\mathcal{A}_{n}(T)$ and $\mathcal{C}_{n}(T)$, are calculated by imposing the boundary conditions (\ref{Bo}), where the latter are developed into a generalized Fourier series of Legendre or Gegenbauer polynomials,
\begin{equation}
    V_{R}\big(\lambda,\theta;T\big)=\sum_{n=1}^{\infty}\Lambda_{n}(T)\mathcal{P}_{n}(\cos\theta),\quad V_{\theta}\big(\lambda,\theta;T\big)=\frac{1}{\sin\theta}\sum_{n=2}^{\infty}\varphi_{n}(T)\mathcal{J}_{n}(\cos\theta).
       \label{fourie}
\end{equation}
The general Fourier coefficients are
\begin{equation}
\begin{aligned}
   \Lambda_{n}(T) &= \cfrac{2n+1}{2}\int_{-1}^{1}{V_{R}\big(\lambda,\xi;T\big)\mathcal{P}_{n}(\xi)\mathrm{d}\xi}\equiv\frac{1}{\lambda^{2}\tilde{\epsilon}^{4}}\bigg[\tilde{\Lambda}_{n}^{(in)}Q_{in}+\tilde{\Lambda}_{n}^{(out)}Q_{out}\bigg]\qquad\\[10pt]
  \varphi_{n}(T) &= \cfrac{n(n-1)(2n-1)}{2}\int_{-1}^{1}{V_{\theta}\big(\lambda,\xi;T\big)\cfrac{\mathcal{J}_{n}(\xi)}{\sqrt{1-\xi^2}}\mathrm{d}\xi}\equiv\frac{1}{\lambda^{2}\tilde{\epsilon}^{4}}\bigg[\tilde{\varphi}_{n}^{(in)}Q_{in}+\tilde{\varphi}_{n}^{(out)}Q_{out}\bigg]
   \label{Lambda}
\end{aligned}
\end{equation}
where
\begin{equation}
\begin{aligned}
    \tilde{\Lambda}_{n}^{(\cdot)}&=\cfrac{2n+1}{2}\int_{I_{(\cdot)}}{ f(\xi;\tilde{\epsilon})\tilde{\omega}(\xi;\tilde{\epsilon}){P}_{n}(\xi)\xi\mathcal\mathrm{d}\xi},
\\[10pt]
    \tilde{\varphi}_{n}^{(\cdot)}&=\cfrac{n(n-1)(2n-1)}{2}\bigg[\cfrac{\Phi^{(\cdot)}\lambda\tilde{\epsilon}^5}{4\sqrt{1-\tilde{\epsilon}^2}}\int_{-1}^{1}{\Theta(\cos^{-1}(\xi))\cfrac{{J}_{n}(\xi)}{\sqrt{1-\xi^2}}\mathcal\mathrm{d}\xi}-\int_{I_{(\cdot)}}{f(\xi;\tilde{\epsilon})\tilde{\omega}(\xi;\tilde{\epsilon}){J}_{n}(\xi)\mathcal\mathrm{d}\xi}\bigg],
   \label{Lambda}
\end{aligned}
\end{equation}
such that $f(\xi;\tilde{\epsilon})=2\big(\xi^2-1+\tilde{\epsilon}^2\big)-\gamma\big(4-\tilde{\epsilon}^2-4\xi^2+3\tilde{\epsilon}^{-2}(\xi^2-1)^2\big)$, the weighed function is $\tilde{\omega}(\xi;\tilde{\epsilon})=\omega\big({S=\tilde{\epsilon}^{-1}\sqrt{1-\xi^2}}\big)$,$\Phi^{(in)}=1$,$\Phi^{(out)}=-1$ ,and the integration's intervals are $I_{(in)}=[-1,-\sqrt{1-\tilde{\epsilon}^{2}}]$ and $I_{(out)}=[\sqrt{1-\tilde{\epsilon}^{2}},1]$. Note that the first integral expression of the $\tilde{\varphi}_{n}$ coefficient in (\ref{Lambda}) is negligible with respect to the second integral expression; therefore, this term can be omitted for simplicity.

\subsection{\textbf{Formulation of the series solution}}
Substitution of (\ref{fourie})-(\ref{Lambda}) into (\ref{GeneralSolution}a-b) where $R=\lambda$ yields two linear equations that define the unknown functions $\mathcal{A}_{n}(T)$ and $\mathcal{C}_{n}(T)$. This provides the solution for the velocity field and the solution is rewritten as,
\begin{subeqnarray}
    V_{R} & = & \cfrac{1}{2}\sum_{n=2}^{\infty}\bigg[
  \big((n+2)\chi^{-2}-n\big)\Lambda_{n-1}+\big(\chi^{2}-1\big)\varphi_{n}\bigg]\chi^{n}\mathcal{P}_{n-1}(\cos\theta),\qquad\qquad\\[10pt]
    V_{\theta} & = & \cfrac{1}{2}\sum_{n=2}^{\infty}\bigg[ 
  \big(\chi^{-2}-1\big)n(n+2)\Lambda_{n-1}+\big(n(1-\chi^{-2})+2\big)\varphi_{n}\bigg]\chi^{n}\cfrac{\mathcal{J}_{n}(\cos\theta)}{\sin\theta}\qquad
\label{LeadingSolution}
\end{subeqnarray}\\
where $\chi(R;T):=R/\lambda(T)\in[0,1]$. The general solution of the pressure distribution can also be rewritten as follows,

\begin{equation}
\label{PressureGeneralSolution}
    P(R,\theta;T)=P_{0}(T)-\varepsilon\cdot\cfrac{1}{\lambda}\sum_{n=1}^{\infty}{\frac{(2n+3)\big((n+1)\Lambda_{n}+\varphi_{n+1}\big)}{n}\chi^{n}\mathcal{P}_{n}(\cos\theta)}.
\end{equation}

This expression represents the pressure distribution inside the chamber, with unknown time-dependent functions. $P_{0}(T)$ may be determined according to the physical boundary conditions of the pressures acting on the flow - at the inlet and outlet. The stretch, $\lambda(T)$, (and therefore also $\Lambda_{n}(\lambda)$ and $\varphi_{n+1}(\lambda)$), may be determined by the wall’s hyperelastic constitutive model. In the next sections, we present two analyses describing different physical cases, including the specific hyperelastic constitutive model we used.

\section{Results}
\label{sec:Results}In this section we obtain the full series solution under various different boundary conditions. First, we present an empirical estimation of $\gamma$ by a problem of flow from an injection tube into a half-space. Then, we present two analyses that describe different physical cases. 
In the first case, we present a chamber whose volumetric flow rate is dictated while the pressure is governed by the chamber's hyperelastic shell. In the second case, we present a chamber whose input pressure is dictated while the wall's hyperelastic law governs the time-varying chamber's radius.
\subsection{\textbf{Estimation of $\gamma$ - Flow from an injection tube into a half-space}}
\label{Estimation}
Assuming the radius of the tube is small relative to the chamber's radius ($\epsilon_{a}\ll 1$), the boundary's polar angle at the connection point is $\pi-\theta_{f}=\sin^{-1}{(\tilde{\epsilon})}= O(\tilde{\epsilon})$ for the inlet tube, and $\theta_{i}=\sin^{-1}{(\tilde{\epsilon})}= O(\tilde{\epsilon})$ for the outlet tube. We focus on the flow field very close to the tube-chamber connection where $\chi \rightarrow 1$ and $\theta \rightarrow \pi$ for inlet tube or $\theta \rightarrow 0$ for outlet tube. Since $\tilde{\epsilon}\ll 1$, it can be modeled as flow from an injection tube into a half-space. 

Several works present numerical analyses for such problems \citep{tutty,weissberg,sisavath,fitz1972}, but there are no analytical solutions to the best of our knowledge. Thus, in order to examine the flow field in the connection region, we utilize finite element schemes devised in COMSOL Multiphysics. In these simulations, the entrapped fluid is modeled according to Navier-Stokes equations, assuming that the flow is incompressible and isothermal. A comprehensive explanation of the numerical simulation will be described in the next part of the results section, while here we restrict ourselves to describing the basic geometry. We assumed a long tube (hundred diameters in length) with unit radius. The boundary condition are non-slip and non-penetration into the tube's edge.

The modified velocity profile we shall use was defined in equation (\ref{ModifiedVelocity}); after setting the non-dimensional variables and neglecting $O(e^{-1/\epsilon_\omega})$ terms, the obtained normalized relation is
\begin{equation}
\label{NDmodefiedVelocity}
    \cfrac{U_{z}^{(m)}}{Q(T)}=2\big(1-S^2\big)+\gamma\big(1-4S^2+3S^4\big).
\end{equation}

\begin{figure}
    \centering
    \includegraphics[width=0.6\textwidth]{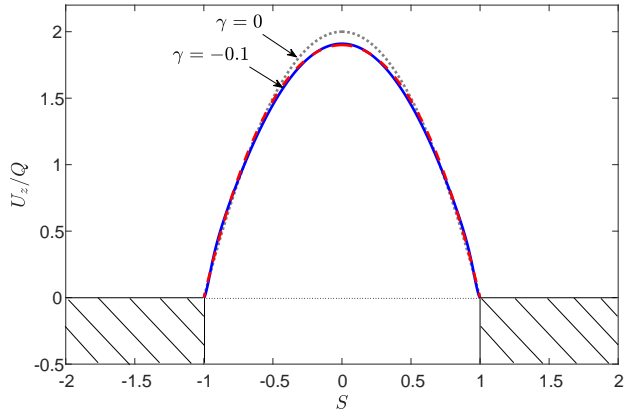}
    \caption{Modified velocity profile (\ref{ModifiedVelocity}) used as a boundary condition of the chamber solution. The gray dotted curve ($\cdot\cdot\cdot$) is the known parabolic Hagen-Poiseuille profile, obtained by setting $\gamma=0$ in the Modified velocity profile (\ref{ModifiedVelocity}). The blue line (\textcolor{blue}{-}) is the velocity profile obtained in the numeric CFD simulation of injection tube into a half-space. The red dashed-line curve (\textcolor{red}{$--$}) is the modified velocity profile (\ref{NDmodefiedVelocity}), with $\gamma=-0.1$ and $\epsilon_{\omega}=0.05$.}
        \label{fig:Boundary}
\end{figure}

Hence, the parameter $\gamma$ should be found by fitting the profile (\ref{NDmodefiedVelocity}) to the numerical simulation results, using the least-squares method. The best fitted value is $\gamma=-0.1$. The velocity profile obtained in the numeric simulation and the modified velocity profile we fitted in (\ref{NDmodefiedVelocity}) is shown in Figure \ref{fig:Boundary}. All the parameters used in the simulations are elaborated in Table \ref{tab:kd}. During the further analysis, we will use the same approximation (with $\gamma=-0.1$).  This approximation will be valid as long as $\tilde{\epsilon}\rightarrow 0$.
\begin{table}
 \begin{center}
\def~{\hphantom{0}}
\begin{tabular}{ |p{5cm}||p{1.5cm}|p{1.5cm}|p{1.5cm}| }
 \hline
 Parameters& Notation & Value & Units\\
 \hline
 Density   & $\rho$   & 1260 &   Kg/m$^{3}$\\
 Dynamic Viscosity&   $\mu$  & 1.1   & Pa$\cdot$s\\
 Elastic Parameter & $s_{1}$ & 1.5 &  MPa\\
 Elastic Parameter & $s_{2}$ & 0.15 &  MPa\\
 Unstressed Radius & $r_{0}$ & 5 &  mm\\
 Unstressed Thickness & $w_{0}$ & 50 &  $\mu$m\\
 Tube radius  & $a$ & 1 &  mm\\
 Tube length  & $\ell$ & 20 &  cm\\ 
 \hline
 \hline
 Volumetric flux  & q & 200 &  mm$^3$/s\\
 Radius  & $\eta$ & 6 &  mm\\ 
 \hline
 \hline
  Tube slenderness  & $\epsilon_{t}$ & $5\times 10^{-3}$ & \\
  Tube-Chamber radii ratio & $\epsilon_{a}$ & $2\times 10^{-1}$ &  \\
  Smoothing parameter & $\epsilon_{\omega}$ & $5\times 10^{-2}$ &  \\ Chamber viscous resistance & $\varepsilon$ & $4\times 10^{-5}$ &  \\
\hline
\end{tabular}
  \caption{Summary of physical parameters values used for plotting the analytical (series) solutions of the dynamic cases. 
  The density and the kinematic viscosity are related to Glycerol which is known as a viscous liquid. The geometric parameters are chosen as the typical value of hyperelastic small chambers used in \cite{BenHaim}.}
  \label{tab:kd}
  \end{center}
\end{table}
\subsection{\textbf{Case I - Dictated inlet flux and hyperelastic wall model}}
\label{Dynamic_case_II}
In this case we dictate the volumetric rate into an elastic chamber. 
The integral mass conservation equation (\ref{integralMass}) is simplified to 
\begin{equation}
\label{lambdaQ}
    \lambda(T)=\bigg[1+\cfrac{3}{4}\int_{0}^{T}\big(Q_{in}(\tau)-Q_{out}(\tau)\big)\mathrm{d}\tau\bigg]^{1/3}.
\end{equation}

While the stretch function $\lambda(T)$ is known, as described in (\ref{lambdaQ}), the inlet pressure is not dictated and additional data regarding the pressure distribution is obtained from the hyper-elastic constitutive relations. Here, it is worth emphasizing that any general constitutive elastic law can serve as a basis for subsequent development, even if it is not bi-stable or hyperelastic.
Integrating the strain density function, $\psi(\lambda)$, over the volume of the chamber's spherical shell and keeping only leading-order terms yields the leading-order chamber's strain energy,
\begin{equation}
\label{Work0}
    \iiint_{\mathbb{V}}{\psi}(\lambda)\mathrm{d}\mathbb{V}=4\pi r_{0}^{2}w_{0}\psi^{*}\cdot\hat{\psi}(\lambda),
\end{equation}
where $\mathbb{V}\approx w\cdot \mathbb{S}$ is the material (constant) volume of the thin shell, $\psi^{*}$ is the characteristic value of $\psi$, and $\hat{\psi}$ is the normalized strain energy density function $\hat{\psi}(\lambda)=\psi(\lambda)/\psi^{*}$.
The work done by the surface traction acting between two states without body force is,
\begin{equation}
\label{Work}
    \iint_{\mathbb{S}}\int_{\xi=r_{0}}^{\eta}\big(p(\xi,\theta)\mathrm{d}\mathbb{\boldsymbol{\mathbb{S}}}\big)\cdot\mathrm{d}\boldsymbol{\xi}=2\pi r_{0}^{3}p^{*}\int_{\theta=\kappa\theta_{i}}^{\theta_{f}}\int_{\xi=1}^{\lambda}\xi^{2}P(\xi,\theta)\sin\theta\mathrm{d}\theta\mathrm{d}\xi.
\end{equation}

Substitution of the pressure's solution (\ref{PressureGeneralSolution}) into (\ref{Work}), and using the mechanical energy principle which states that the work done by the surface tractions acting between two equilibrium states without body force is balanced by the change in the total strain energy \citep{Beatty}, allow us to fully define the pressure distribution in the chamber,
\begin{equation}
\label{PressureCase2}
  \begin{split}
    P_{(I)}(R,\theta;T)&=\frac{1}{\lambda^2}\cfrac{d\hat{\psi}}{d\lambda}-\\[10pt]
    &-\varepsilon\cdot\frac{1}{\lambda}\sum_{n=1}^{\infty}{\frac{(2n+3)\big((n+1)\Lambda_{n}+\varphi_{n+1}\big)}{n}\bigg[\cfrac{1}{2}\big(\mathbb{P}_{n}^{(in)}+\kappa\mathbb{P}_{n}^{(out)}\big)+\chi^{n}\mathcal{P}_{n}(\cos\theta)}\bigg],
\end{split}  
\end{equation}
where $\mathbb{P}_{n}^{(\cdot)}$ is the zero$^{th}$ moment of $\mathcal{P}_{n}(\xi)$ about an origin, defined as
\begin{subeqnarray}
    \mathbb{P}_{n}^{(in)}&\equiv&\int_{-1}^{-\sqrt{1-\tilde{\epsilon}^{2}}}{\mathcal{P}_{n}(\xi)\mathrm{d}\xi}  =  \cfrac{\mathcal{P}_{n+1}\big(-\sqrt{1-\tilde{\epsilon}^{2}}\big)-\mathcal{P}_{n-1}\big(-\sqrt{1-\tilde{\epsilon}^{2}}\big)}{2n+1},\\[10pt]
    \mathbb{P}_{n}^{(out)}&\equiv&\int_{\sqrt{1-\tilde{\epsilon}^{2}}}^{1}{\mathcal{P}_{n}(\xi)\mathrm{d}\xi}  =  -\cfrac{\mathcal{P}_{n+1}\big(\sqrt{1-\tilde{\epsilon}^{2}}\big)-\mathcal{P}_{n-1}\big(\sqrt{1-\tilde{\epsilon}^{2}}\big)}{2n+1}.
\label{P_int}
\end{subeqnarray}

From order of magnitude analysis in (\ref{Work0}) and (\ref{Work}), we obtain the characteristic pressure, which depends on the specific hyper-elastic model we use,
\begin{equation}
    p^{*}=\cfrac{w_{0}}{r_{0}}\psi^{*}.
\end{equation}

According to the solution obtained in (\ref{PressureCase2}), the pressure distribution inside the chamber consists of two parts. The first part is the well-known isotropic pressure obtained by \cite{Beatty}. This expression represents the isotropic pressure in the leading order, which is experienced by the chamber's elastic wall, assuming that the pressure is uniform and equal to $P_{S}(\lambda)=\lambda^{-2}\cdot d\psi/d\lambda$.  The second expression is the transient developing pressure profile, $\varepsilon P_{1}(R,\theta;\tilde{\epsilon};T)$, which varies spatially and temporally.

\subsubsection{Numerical verification of the fully coupled model}

\begin{figure}
  \centering
    \includegraphics[width=1\textwidth]{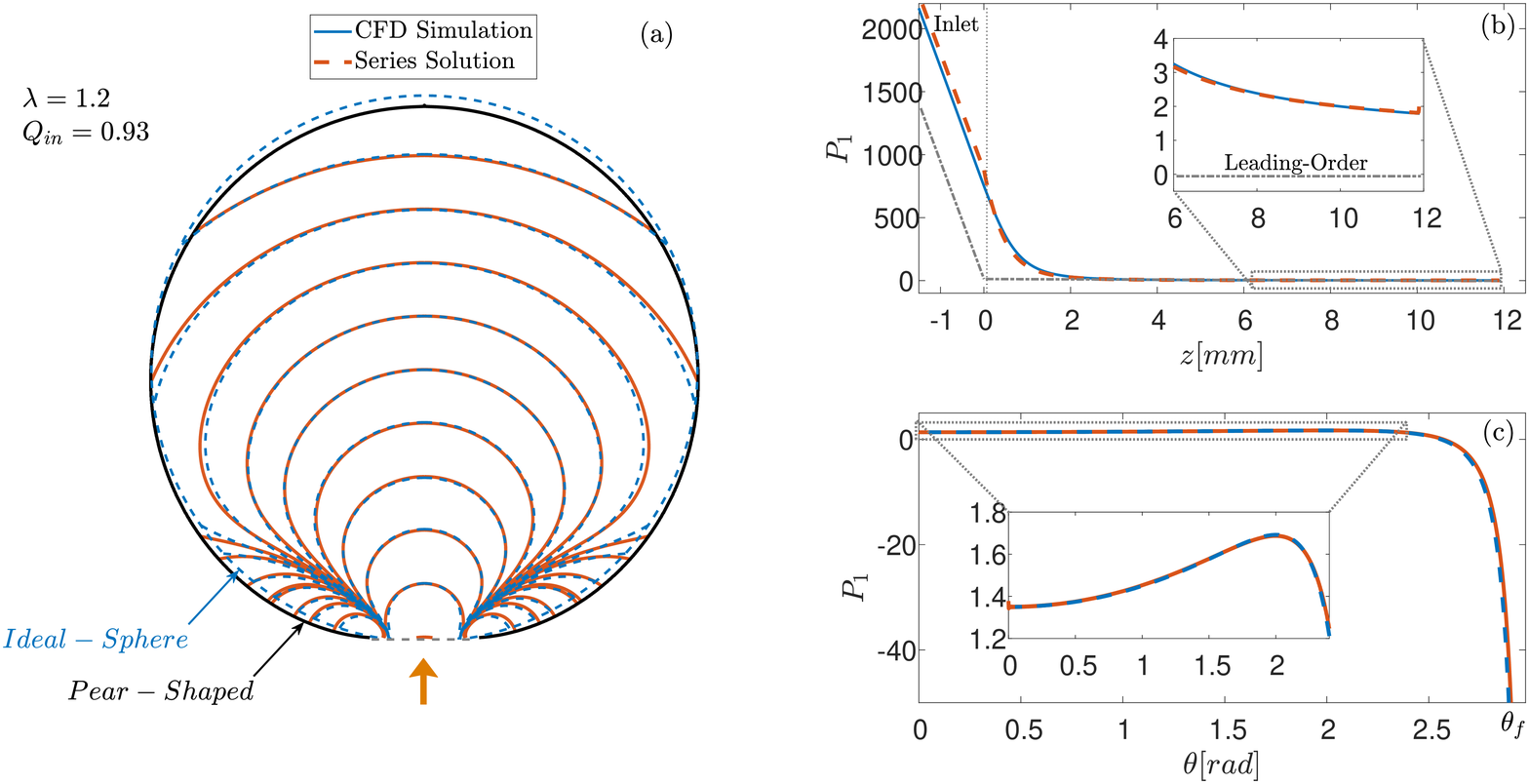}
     \includegraphics[width=1.0\textwidth]{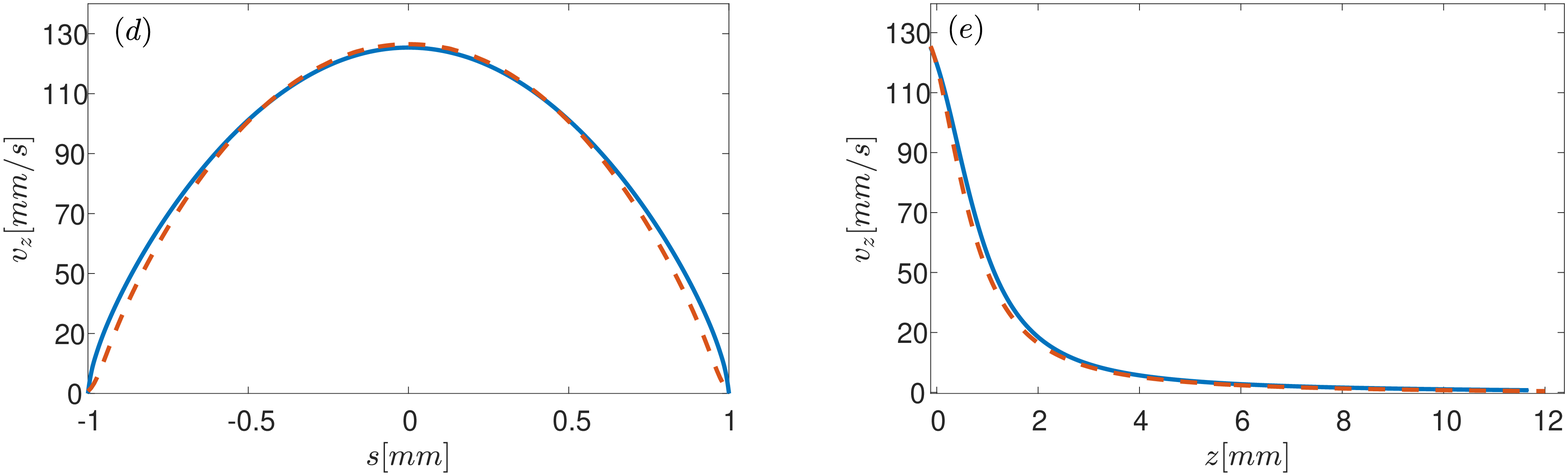}
         \caption{A numerical verification of the fully coupled model. The chamber's radius is $6mm$, and the volumetric flux is $q=200mm^3/s$. The dashed red line (\textcolor{red}{- - -}) is obtained by the analytical (series) solution, which was calculated based on the first 100 terms in the series. The blue line (\textcolor{blue}{---}) is obtained in the CFD simulation. Excellent agreement is observed. (a) Contour curves of the pressure distribution obtained from the simulation (red curves) and the same contour curves obtained from the series solution (dashed blue curves). The black non-spherical (pear-shaped) boundary describes the elastic chamber wall obtained from the simulation. (b) The non-dimensional dynamic pressure along the symmetry axis $z$. (c) The non-dimensional dynamic pressure at the chamber wall as a function of the angle $\theta$.(d) The axial velocity of the flow relative to a cylindrical coordinate system. (e) The axial velocity component of the flow along the axis of symmetry.}
    \label{fig:SingleInputVelocitiesCFD}
\end{figure}
In order to validate the theoretical model, we have utilized commercially available software (COMSOL multiphysics) in order to conduct finite element simulations considering the fully-coupled dynamics of the system. In these simulations, the entrapped fluid is modeled according to Navier-Stokes equations, assuming that the flow is incompressible and isothermal. Moreover, the shell is modeled according to the Mooney-Rivlin model, wherein contrary to the theoretical model, it is not restricted to a spherical shape. All the parameters used in the simulations are elaborated in Table \ref{tab:kd}.
In all simulations, the geometry and boundary conditions are taken in correspondence with the problem statement in section § \ref{sec:Problem_Formulation}. Further, the physics is described by the fluid-structure interaction module of COMSOL, referring to the chamber as a two-parameter Mooney-Rivlin hyperelastic solid. This is done while employing a moving mesh formulation to accommodate the changes of the fluid's domain. The coupling between the solid and the fluid is carried out by balancing the fluid's velocity and the time derivative of the solid's displacements and the normal components of their stress tensors at the interface between the solid and the fluid. The numerical schemes describe the deformation field of the solid utilizing second-order base functions, where those used to discretize the velocity field and pressure distribution of the fluid are cubic and quartic, respectively. Finally, the system's geometry is described as two-dimensional and axisymmetric to eliminate significant numerical errors and decrease the computational effort.
Furthermore, the meshing of the geometry was enhanced until low sensitivity to further refinement was achieved. In the final meshing used in the simulations whose results are presented here, the solid is modeled by rectangular elements whose grid contains 200 tangential elements and six elements along with its thickness. Similarly, the tube is modeled by a regular grid having 30 radial and 103 axial elements. Finally, since the geometry of the fluid residing inside the chamber is of higher complexity, the latter is described by free triangular elements. The maximal size of these elements is restricted to 0.18 mm. On the boundaries of this region, where the sensitivity is higher, and the precision is of greater importance, the maximal size is further reduced to 0.1 mm. 

While numerically simulating the dynamics of the fluid-filled chamber, utilizing the two constant parameters of the Moony-Rivlin model used in the theoretical computations, we noticed discrepancies. By examining the numerically computed static pressure-stretch relation, it was apparent that these discrepancies might originate from errors in the hyperelastic module of COMSOL Multiphysics version 5.0, in which all the numerical simulations were carried out. Therefore, to compare the theoretical and numerically simulated dynamic responses of the fully coupled system, the hyperelastic model used in the numerical scheme had to be modified. To calibrate this model, we fitted the pressure-stretch relation of the numerical scheme at a static regime. Namely, we found the appropriate values of the two Moony-Rivlin constant parameters, leading to the desired theoretical static pressure-stretch relation while inflating the chamber to different radii, corresponding to different stretches. The values of the pressure inside the chamber and its effective radii were taken a long period after the inflation was finished after all dynamic effects decayed, including the non-spherical modes. As a result, the values used in the simulations yield a relative error of less than $0.4\%$ in the pressure-stretch relation at the presented values.

Figure \ref{fig:SingleInputVelocitiesCFD}, which compares the theoretical and the numerically-simulated pressure-distribution, shows an excellent agreement, thus validating the analytical model and its underlying assumptions. In addition, a numerical investigation of the pressure distribution where $\chi\rightarrow 1$ (inner chamber's wall region) displayed that the varies spatially and temporally correction is $O(\epsilon_{t})$. This result is consistent with the analytical result we obtained in the next section.

\subsubsection{Additional dynamic cases}
\begin{figure}
    \centering
    \includegraphics[width=1.0\textwidth]{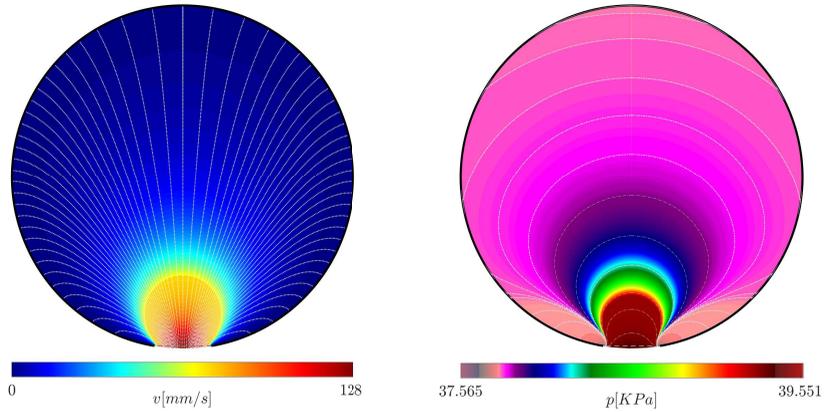}
    \caption{Series solution of single inlet chamber (Case I): Left - velocity field and stream-lines, where the colors describe the magnitude of the flow velocity, and the white dotted lines $(\cdot\cdot\cdot)$ are the stream-lines. Right - pressure distribution with white dash-dotted $(-\cdot-)$ contour lines of constant pressure.  The chamber's radius is $6$ mm and the volumetric flux is $q=200$ mm$^3$/s .\\[5pt]}
    \label{fig:SingleInput}
    \end{figure}
    \begin{figure}
  \centering
    \includegraphics[width=1.0\textwidth]{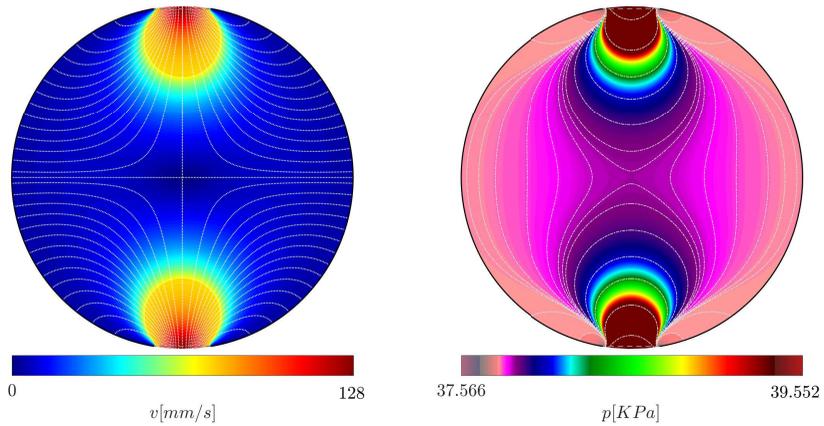}
    \caption{Series solution of a double inlet chamber (Case I): left - velocity field magnitude (color-map) and stream-lines (white dotted curves), right - pressure distribution. The chamber's radius is $6mm$ and the volumetric flux is $q=200mm^3/s$ in both inlets. In this case, each streamline is directed to the chamber's wall.}
    \label{fig:DoubleInput}
\end{figure}
\begin{figure}
  \centering
    \includegraphics[width=1.0\textwidth]{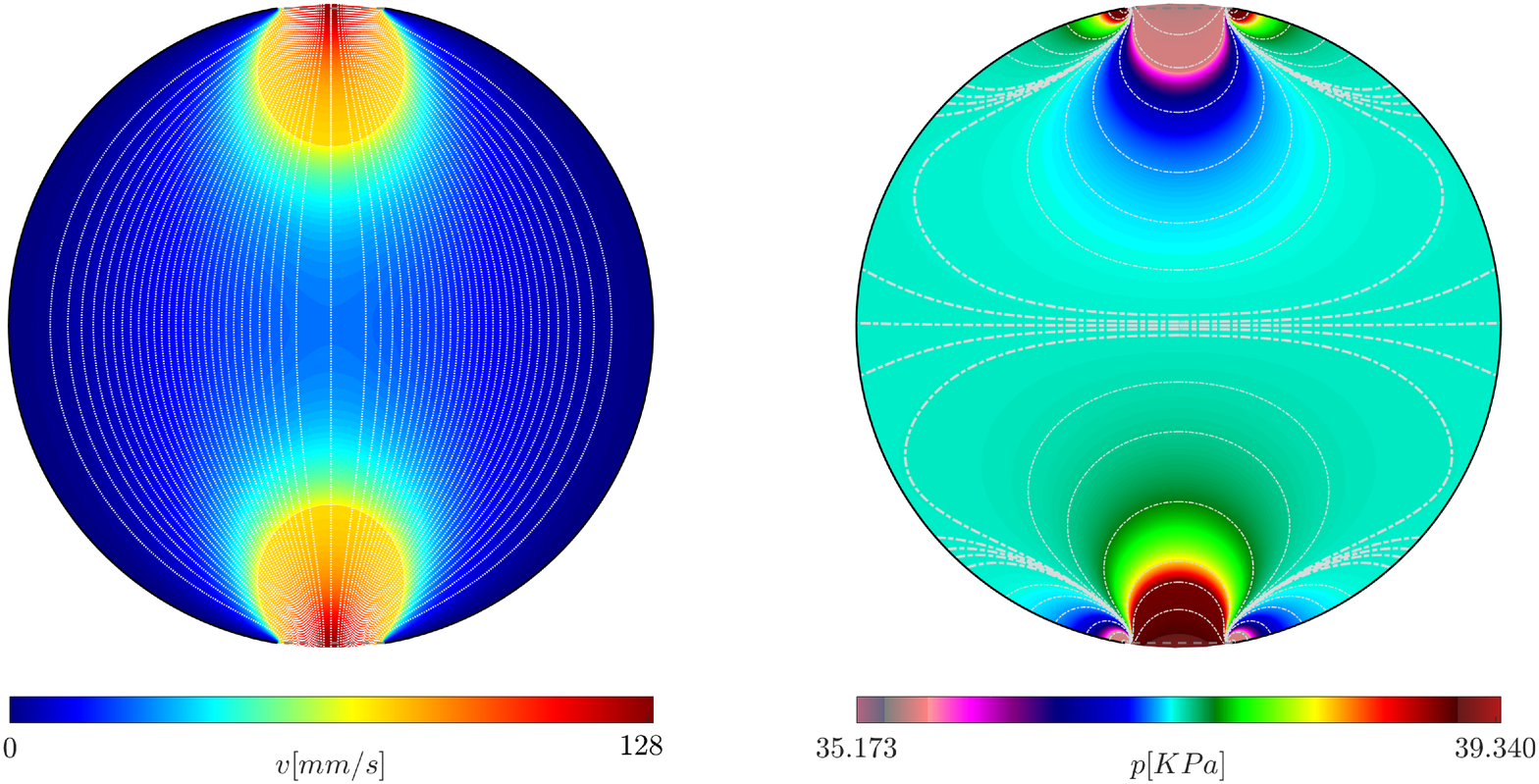}
    \caption{Series solution of a chamber with inlet and outlet (Case I): left - velocity field magnitude (color-map) and stream-lines (white dotted curves), right - pressure distribution. The chamber's radius is $6mm$ and the volumetric flux is $q=200mm^3/s$ in both inlet and outlet. Since the flow rate of the inlet and outlet are equal, a steady-state solution is obtained for the flow and pressure fields. After an initial transient, the flow and pressure fields reach a steady state.}
    \label{fig:InputOutput}
\end{figure}
Figure \ref{fig:SingleInput} presents the series solution of the flow velocity magnitude with streamlines and the pressure distribution, based on equations (\ref{LeadingSolution}) - (\ref{PressureCase2}), and utilizing the parameters in Table \ref{tab:kd}. The expected characteristic pressure is $p^{*}=$ O(1KPa), the characteristic flow velocity within the chamber is $v^{*}=$ O(1mm/s) and $Re\approx0.01$. The results are based on summation of $100$ terms in the series solution in (\ref{PressureCase2}). A remarkable and non-intuitive result was obtained from the pressure solution on the chamber's wall (see Figure \ref{fig:SingleInputVelocitiesCFD}c). The maximum pressure obtained on the elastic wall is not obtained near the inlet tube. This result may be critical in identifying the failure point of the chamber’s wall.

The theoretical solution also readily allows analyzing chambers with two inlets, both with controlled volumetric flow rates. In the first case, whose typical flow velocity magnitude and pressure distributions are presented in Figure \ref{fig:DoubleInput}, the chamber is inflated by equal flow rates from both inlets, whereas in the second case, whose behavior is presented in Figure \ref{fig:InputOutput}, the flow rate in one of the inlets is reversed, meaning that the chamber is inflated and deflated simultaneously. In the first case, each streamline is directed to the chamber's wall; thus, the inflation rate is maximal. 
In the second case, even though the rate of the inlet and outlet are equal, and thus the volume are constant, a pressure distribution is developed that varies in space and not in time.\\ 
\subsection{\textbf{Case II - Dictated inlet pressure and hyper-elastic wall model} }
\label{Dynamic_case_III}
In this case, we dictate only the inlet pressure, and solve for the fluidic pressure distribution, as well as the chamber's stretch, $\lambda(T)$.
We use  integral mass conservation \ref{integralMass}) in order to calculate the pressure at the location in which  the tube connects with the chamber, 
\begin{equation}
    P_{C}(T)\approx P_{in}(T)-32\lambda^2\cfrac{d\lambda}{dT}.
\end{equation}

Similarly to the previous case, we substitute $(R,\theta)=(\lambda,\pi)$ in the pressure distribution solution (\ref{PressureGeneralSolution}c) and equate it to $P_{C}(T)$ in order to solve for the function $P_{0}(T)$. The pressure distribution obtained by these means is
\begin{equation}
\label{Pcase3}
  \begin{split}
    P_{(II)}(R,\theta;T)&=P_{in}(T)-32\lambda^{2}\cfrac{d\lambda}{dT}+\\[10pt]
    &+\frac{4\varepsilon}{\lambda\tilde{\epsilon}^{4}}\frac{d\lambda}{dT}\sum_{n=1}^{\infty}{\cfrac{(2n+3)\big((n+1)\tilde{\Lambda}_{n}^{(in)}+\tilde{\varphi}_{n+1}^{(in)}\big)}{n}\bigg[(-1)^{n}-\chi^{n}\mathcal{P}_{n}(\cos\theta)}\bigg].
\end{split}      
\end{equation}

Finally, using  the mechanical energy principle (\ref{Work}), we derive a nonlinear ordinary differential equation governing  the time-dependent stretch of the chamber,
\begin{equation}
\label{ODE_lambda}
    \bigg[32\lambda^2-\frac{2\varepsilon}{\lambda\tilde{\epsilon}^{4}}\cdot\Upsilon(\tilde{\epsilon})\bigg]\frac{d\lambda}{dT}=P_{in}(T)-\frac{1}{\lambda^{2}}\frac{d\hat{\psi}}{d\lambda},
\end{equation}
where
\begin{equation}
\label{UpsilonSeries}
    \Upsilon(\tilde{\epsilon}):=\sum_{n=1}^{\infty}\frac{(2n+3)\big((n+1)\tilde{\Lambda}_{n}^{(in)}+\tilde{\varphi}_{n+1}^{(in)}\big)}{n}\bigg[2(-1)^n+\mathbb{P}_{n}^{(in)}\bigg]=O(\tilde{\epsilon}).
\end{equation}

The function $\Upsilon(\tilde{\epsilon})$ was estimated numerically by summing the first $10^5$ terms in the series for several values of $\tilde{\epsilon}$ corresponding to $\tilde{\epsilon}\in[0.002,0.5]$.
By substituting the asymptotic approximation of $\Upsilon$, into the differential equation that governs the chamber's stretch (\ref{ODE_lambda}), we obtain its approximated explicit form given by
\begin{equation}
\label{ODEsimplified}
    \cfrac{d\lambda}{d T}=\cfrac{1}{32\lambda^{2}}\bigg[P_{in}(T)-\frac{1}{\lambda^{2}}\frac{d\hat{\psi}}{d\lambda}\bigg]+O(\epsilon_{t}).
\end{equation}

\begin{figure}
  \centering
    \includegraphics[width=1.0\textwidth]{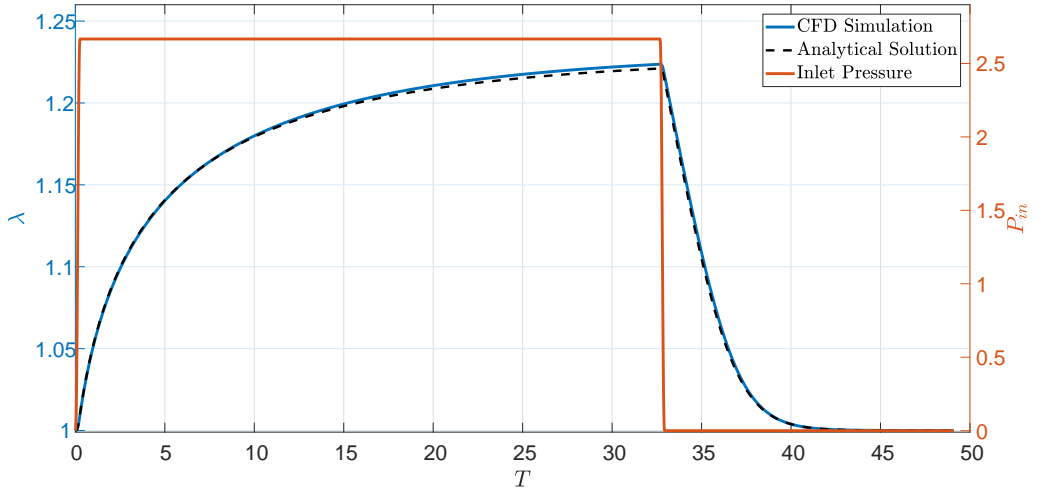}
    \caption{A numerical verification of the fully coupled model describing a chamber's dynamic responses to a pressure pulse imposed at a single inlet, as shown in the red line. The continuous blue curve represents the solution obtained in the numerical CFD simulation, and the dashed black line represents the solution obtained by the equation (\ref{ODEsimplified}), developed in our analysis. An excellent match between the results can be observed.}
    \label{fig:StretchVaringCFD}
\end{figure}

Next, we investigate the system's behavior in (\ref{ODEsimplified}), whose motion is governed by a controlled pressure inlet. In order to validate this model, we have compared its solution obtained utilizing the parameters in table \ref{tab:kd} to finite element simulations carried out in COMSOL Multiphysics. The numerical scheme used here is similar to the one utilized in the previous section, but with the imposed flow rate replaced with an imposed piecewise constant pressure.
Figure \ref{fig:StretchVaringCFD} compares the stretch of the chamber in time, as achieved theoretically from the asymptotic equation (\ref{ODEsimplified}) and numerically from the simulation. Since the chamber is not restricted to be spherical in the numerical simulations, the stretch is taken to be its effective value given in \cite{Ilssar} as $\eta_{eff}(t)=\sqrt{A_{S}(t)/4\pi}$. Here, $A_{S}(t)$ is the body's surface area obtained in the non-spherical simulations, and $\eta_{eff}(t)$ is the effective radius of an ideal sphere having the same surface area as the non-spherical body. This figure shows an excellent agreement.

In order to approximate the characteristic time constant of the system, enabling us to estimate the period it takes to reach a steady-state, we formulate a linear approximation. The linear system describing the system’s dynamic response, close to an equilibrium point given by $(\lambda_{SS},P_{SS})$, when the pressure at the inlet is dictated to be $P_{ext}(\tau)$. The linear equation is solved analytically, leading to the following solution:
\begin{equation}
\label{linearapprox}
    \lambda_{L}(\tau)=\lambda_{SS}+\bigg(\lambda(0)-\lambda_{SS}+\frac{1}{4\lambda_{SS}^{2}}\int_{0}^{\tau}{e^{\beta_{I}\cdot\tau'}\triangle P_{in}(\tau')}d\tau'\bigg)e^{-\beta_{I}\cdot\tau}.
\end{equation}  
where $\triangle\lambda_{L}=\lambda_{L}(\tau)-\lambda_{SS}$ is a small stretch variation around its nominal value $\lambda_{L}$, $\triangle P_{in}(\tau)=P_{in}(\tau)-P_{SS}$ is the pressure variation from its nominal value, and
\begin{equation}
\label{betaa}
    \beta_{I}=\frac{1}{4\lambda_{SS}^{2}}\frac{dP_{SS}}{d\lambda_{SS}}=\frac{P_{SS}}{2\lambda^{3}_{SS}}-\frac{3}{\lambda_{SS}^{4}}+\frac{9}{\lambda_{SS}^{10}}-\alpha\bigg(\frac{1}{\lambda_{SS}^{2}}-\frac{7}{\lambda_{SS}^{8}}\bigg).
\end{equation}

From (\ref{betaa}), it is clear that the solution branches of $P_{SS}$ in (\ref{Pss}) are stable equilibria if and only if $dP_{SS}/d\lambda_{SS}>0$. The stability criterion obtained from the equation's linearization is identical to the stability criterion obtained from energetic considerations in (\ref{second_derivative}). Moreover, from the linear solution, the relaxation time can be approximated as $T_{relax}=32/\beta_{I}$.
Since the derivative $dP_{SS}/d\lambda_{SS}$ in the first branch (I) of the typical pressure-stretch curve, which is plotted in Figure. {\ref{fig:staticCase}}, is significantly higher than in the third branch (III), 
the dynamic response in the third region is much slower.

In order to achieve a better approximation, equation (\ref{ODEsimplified}) is approximated by a quadratic Taylor expansion around the general equilibrium point. When the pressure at the inlet equals the steady-state pressure, $\triangle P_{in}(\tau)=0$, the solution of this equation under small initial perturbation from equilibrium $\lambda_{Q}(T)$ is given by:
\begin{equation}
\label{secondorderapprox}
    \lambda_{Q}(\tau)=\lambda_{SS}+\cfrac{\beta_{I}}{-\beta_{II}+\bigg(\beta_{II}+\cfrac{\beta_{I}}{\lambda(0)-\lambda_{SS}}\bigg)e^{\beta_{I}\tau}}
\end{equation}
where
\begin{equation}
    \beta_{II}=-\frac{3P_{SS}}{4\lambda^{4}_{SS}}+\frac{6}{\lambda_{SS}^{5}}-\frac{45}{\lambda_{SS}^{11}}+\alpha\bigg(\frac{1}{\lambda_{SS}^{3}}-\frac{28}{\lambda_{SS}^{9}}\bigg).
\end{equation}
\begin{figure}
    \centering
    \includegraphics[width=1.1\textwidth]{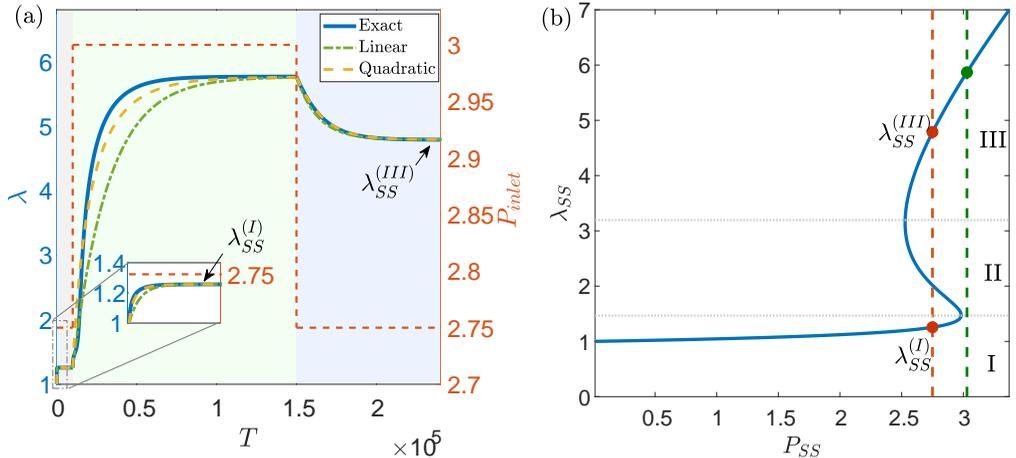}
    \caption{Analytical Solutions: (a) - The evolution of the chamber's stretch (solid blue curve), alongside its linear approximation (\ref{linearapprox}) (dash-dot green curve), and quadratic approximation (\ref{secondorderapprox}) (dashed yellow curve) , where the input pressure (dashed red curve) varies between its extremum values $P_{A}$ and $P_{B}$. 
    (b) - The equilibrium pressure-stretch curve (solid blue curve), alongside the lower (dashed red line) and higher (dashed green line) pressure values, corresponding to $P_{A}$ and $P_{B}$. The red dashed line shows the entry pressure value in the first and last stage ($P_{SS}=2.75$), and the green dashed line shows the pressure dictated in the middle stage ($P_{SS}=3.05$). The equilibrium points corresponding to the $P_{SS}=2.75$ pressure are marked with red marks where $\lambda_{SS}^{(I)}=1.25$ and $\lambda_{SS}^{(III)}=4.79$.}
    \label{fig:bistable}
\end{figure}

In Figure (\ref{fig:bistable}), the linear and the second-order approximations were displayed alongside the exact numerical solution of equation (\ref{ODEsimplified}). The excellent agreement between the exact dynamic response and both approximations indicates that these two approximations are suitable for the prediction of the chamber's evolution.

In section \ref{ElsticSection}, we have shown that in the pressure range spanning between $P_{A}$ and $P_{B}$, there are three possible equilibrium radii for each constant pressure value. Multiple solutions can be exploited to switch from one equilibrium state to another under the same steady-state pressure while passing through the unstable, middle branch.
An example of such a transition is shown in Figure \ref{fig:bistable}, where after increasing the input pressure and decreasing it back to its initial value, the chamber's radius does not return to its initial radius. Instead, it retains a larger radius, corresponding to the higher equilibrium state. Initially, the system converges to the equilibrium point $\lambda_{SS}^{I}$ corresponding to $P_{SS}$, in the first branch (I) which is closer to the initial conditions; then, the inlet pressure rises to a value higher than $P_{A}$ to move the chamber to another equilibrium point in the third branch (III). Finally, we decrease the pressure again to the same level as the initial step, $P_{SS}$, so that the system will converge to the second equilibrium point in the third branch (III). This results are consistent with the insights raised in our previous work \citep{BenHaim}.
It is worth emphasizing at the end of this section that the general solutions obtained in relations (\ref{LeadingSolution}),(\ref{PressureCase2}) and (\ref{Pcase3}), are parametrically dependent on the dictated strain energy density function  $\psi(\lambda)$, based on the chosen constitutive law. Here, we have chosen to present the results using Mooney-Rivlin's hyperelastic constitutive law to illustrate the phenomenon of bi-stability and to compare our results with some other works which have assumed uniform pressure. However, any other constitutive law for the elastic shell can be used in order to derive the solution for the pressure distribution and velocity field in the three cases illustrated in the section.\\
\section{\textbf{The dynamic behavior of two interconnected bi-stable chambers}}
\label{TwoBalloons} 
In this section, we analyze the behavior of a system consisting of two chambers, serially connected through slender tubes to a single inlet, whose flow rate is dictated and equal to $Q_{in}(T)\equiv Q(T)$. This system is instrumental for understanding the behavior of interconnected bi-stable elements, and it sheds light on the capability to govern the constituent elements by employing a single input \citep{BenHaim}. In many works in which flow-controlled bi-stable elastic systems have been examined, the main assumption is uniform pressure within the elastic element  \citep{BenHaim,Dreyer,Treloar,Glozman}. Here, we shall analyze the dynamics of such systems using the solution we developed in the previous sections, which considers the pressure distribution in the elastic element's inner space. This system's physics is described utilizing the analyses presented above, where  we study identical tubes and chambers. The system under investigation combines two of the cases analyzed in the previous section, as the flow from the inlet to the first chamber is dictated. 
We denote the stretch of the first and the second chambers as $\lambda_{1}(T),\lambda_{2}(T)$, respectively. From the integral mass rate balance, 
\begin{equation}
\label{massrate}
    Q(T)=4\lambda_{1}^2\cfrac{d\lambda_{1}}{dT}+4\lambda_{2}^2\cfrac{d\lambda_{2}}{dT}.
\end{equation}

The stretch $\lambda_{2}(T)$ of the second chamber is governed by (\ref{ODEsimplified}), where $P_{in}(T)$ is related to the pressure in the point in the tube relative to the connection with the (second) chamber. In our case $P_{in}^{(eff)}(T)=P(\lambda_{1},0;T)$ is the effective external pressure, where $P(R,\theta;T)$ is given by (\ref{PressureCase2}). Recalling that the Fourier coefficients $\Lambda_{n}$ and $\varphi_{n}$ are linear combinations of the inlet the outlet flow rates (\ref{Lambda}), the effective external pressure is rewritten as 
\begin{equation}
\label{PressureEffective}
    P_{in}^{(eff)}(T) = \cfrac{1}{\lambda_{1}^{2}}\frac{d\hat{\psi}}{d\lambda}\bigg|_{\lambda_{1}}+\frac{\varepsilon}{\lambda_{1}^{3}\tilde{\epsilon}_{1}^{4}}\bigg(4\lambda_{2}^{2}\frac{d\lambda_{2}}{dT}\cdot\Pi^{(out)}(\tilde{\epsilon}_{1})+Q(T)\cdot\Pi^{(in)}(\tilde{\epsilon}_{1}) \bigg),
\end{equation}
where
\begin{equation}
\label{PiFunction}
    \Pi^{(\cdot)}(\tilde{\epsilon}_{1}):=\sum_{n=1}^{\infty}\cfrac{(2n+3)\big((n+1)\tilde{\Lambda}_{n}^{(\cdot)}+\tilde{\varphi}_{n+1}^{(\cdot)}\big)}{n}\bigg[-\frac{1}{2}\big(\mathbb{P}_{n}^{(in)}+\mathbb{P}_{n}^{(out)}\big)-1\bigg],
\end{equation}
and $\tilde{\epsilon}_{1}(T)=\epsilon/\lambda_{1}(T)$.
\begin{figure}
    \centering
    \includegraphics[width=1.0\textwidth]{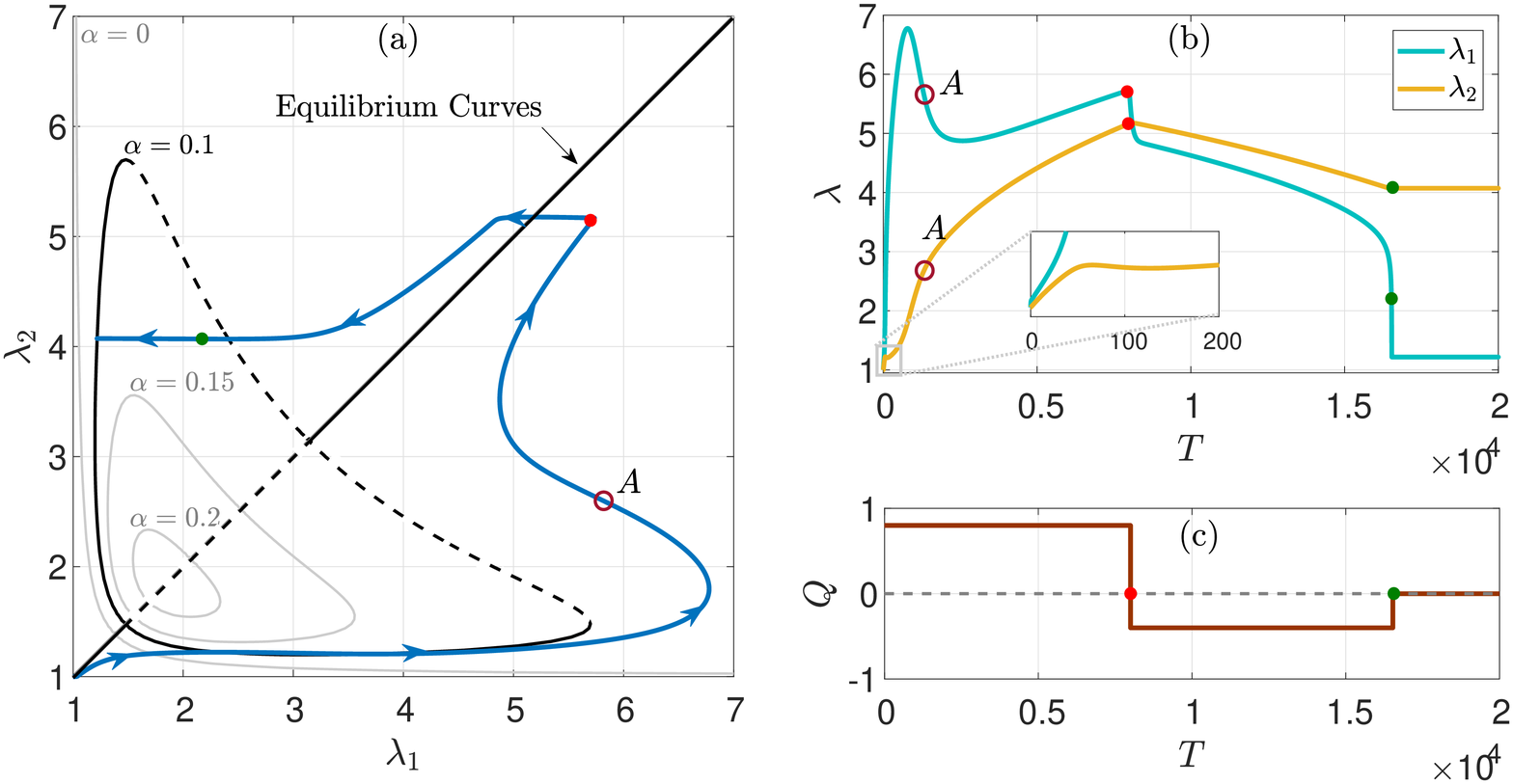}
    \caption{(a) - Equilibrium curves of the two-chamber system in $\{\lambda_{1};\lambda_{2}\}$ plane. Black solid curves are stable branches, and black dashed curves are unstable ones for $\alpha=0.1$. The evolution of the equilibrium curves by the $\alpha$ parameter is described in the grey curves. The blue line is the solution trajectories of numerical simulations of (\ref{massrate}) and (\ref{ODEtwo}), overlaid on the branches of equilibrium curves. The red and green points describe the moments in which the flux is changed. (b) Time plots of chambers’ stretch $\lambda_{i}(T)$ obtained by numerical integration of the nonlinear dynamical system. (c) - Time plot of inlet flow $Q$ for inflation and deflation in the case of two chambers.}
    \label{fig:ODE45}
\end{figure}
\begin{figure}
    \centering
    \includegraphics[width=1.0\textwidth]{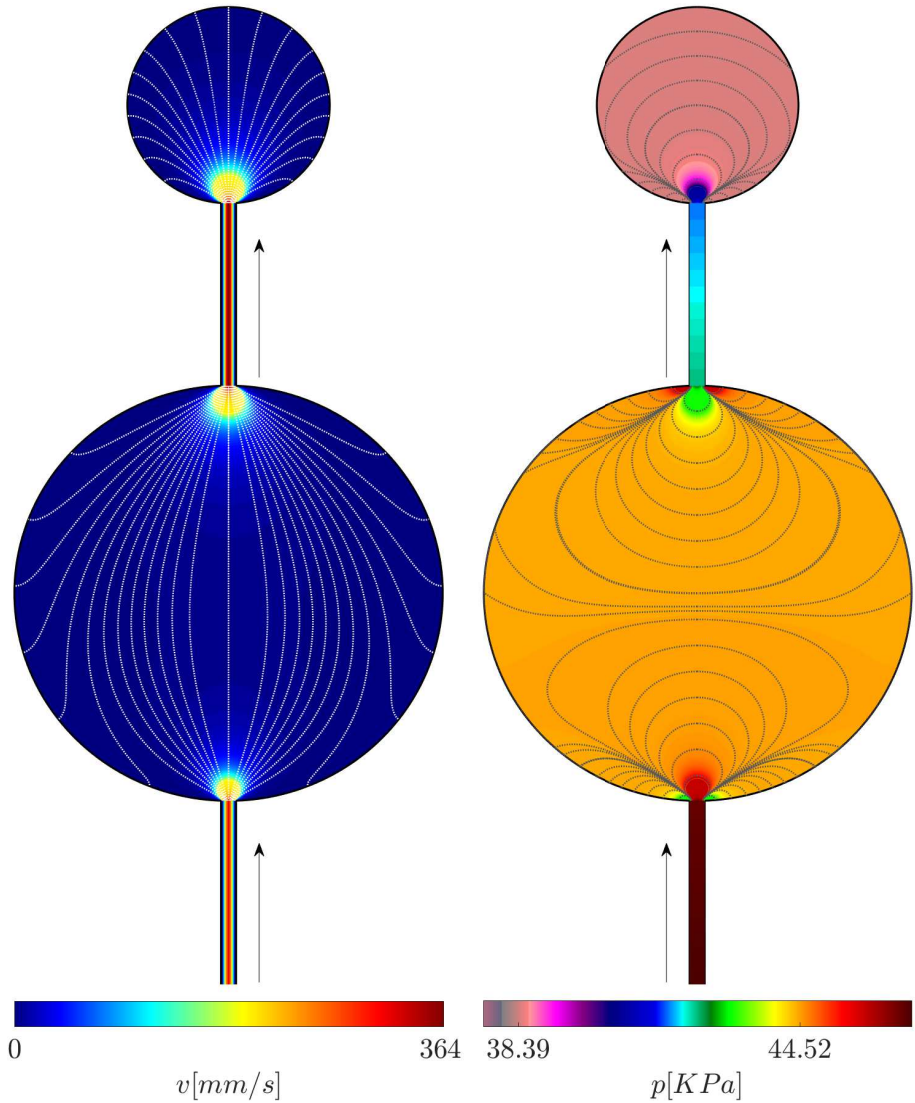}
    \caption{Series solution of a system consisting of two chambers controlled by single inlet: Left - velocity field and stream-lines, Right - pressure distribution. The flow velocity and pressure distribution corresponding to when the system passes through the point A shown in Figure \ref{fig:ODE45}.}
    \label{fig:TwoChamber}
\end{figure}
Utilizing the results achieved for the different values of $\tilde{\epsilon}_{1}$ yields the following approximations $\Pi^{(out)}(\tilde{\epsilon}_{1})=O(\tilde{\epsilon}_{1})$ and $\Pi^{(in)}(\tilde{\epsilon}_{1})=O(\tilde{\epsilon}_{1}^{2})$.
Using the governing equation of $\lambda_{2}(T)$ in (\ref{ODEsimplified}) yields the second equation of motion,
\begin{equation}
\label{ODEtwo}
    \cfrac{d\lambda_{2}}{dT}=\cfrac{1}{32\lambda_{2}^{2}}\bigg(\frac{1}{\lambda_{1}^{2}}\frac{d\hat{\psi}}{d\lambda}\bigg|_{\lambda_{1}}-\frac{1}{\lambda_{2}^{2}}\frac{d\hat{\psi}}{d\lambda}\bigg|_{\lambda_{2}}\bigg)+O(\epsilon_t).
\end{equation}

Equations (\ref{massrate}) and (\ref{ODEtwo}) are a set of nonlinear coupled first-order differential equations that govern the evolution of the chambers’ stretches $\lambda_{i}(T)$ under the single input $Q(T)$. In our previous work \citep{BenHaim}, we have presented an algorithm whose purpose is to bring the system from one equilibrium state to another by a single input. There, it is assumed that the process is quasi-static; thus, the chamber's pressure is uniform during the process. Thanks to the analysis presented in this work, it is possible to consider the pressure distribution and the chambers' flow field during the dynamic process.
The equilibrium state of this system is achieved when the flow rate is zero. In this case, the derivatives in time are equal to zero, and the differential equations degenerate into a single algebraic equation that defines the equilibrium curves. The equilibrium curves are defined by,
\begin{equation}
\label{Equilibrium}
   \bigg(\cfrac{1}{\lambda_{1}^{2}}\cfrac{d\hat{\psi}}{d\lambda}\bigg)\bigg|_{\lambda_{1,SS}} = \bigg(\cfrac{1}{\lambda_{2}^{2}}\cfrac{d\hat{\psi}}{d\lambda}\bigg)\bigg|_{\lambda_{2,SS}}.
\end{equation}

Equation (\ref{Equilibrium}) describes the equilibrium curves of the system presented in Figure \ref{fig:ODE45}(a). 
Importantly, this nonlinear equation gives rise to two solutions. One solution is given by $\lambda_{1,SS}=\lambda_{2,SS}$, where the radii of both chambers are equal, whereas in the second one the radii are different, $\lambda_{1,SS}\neq\lambda_{2,SS}$, thanks to the bi-stability of the chambers. When the system is initially placed out of equilibrium (by setting zero input, or response to initial condition), the solution moves along the curve of constant total volume,  $\lambda_{1}^{3}+\lambda_{2}^{3}=\text{Const}$, and converges toward stable equilibrium branches. 
In our previous work \citep{BenHaim},  we have presented an algorithm whose purpose is to bring the system from one equilibrium state to another using a single input. It assumed that the process is quasi-static; thus, the chamber's pressure is hydrostatic during the process. Thanks to the analysis presented in this work, it is possible to consider the pressure profile and the chambers' flow field during the dynamic process. In Figure \ref{fig:ODE45}, we used our algorithm and presented a scenario where the system undergoes an irreversible sequence of transitions between the chambers' combined states while being controlled by a single input of flow rate $Q(T)$. The chosen input is piecewise constant, represented in \ref{fig:ODE45}(c). Figure \ref{fig:ODE45}(b) shows the system’s trajectory in $\{\lambda_{2},\lambda_{2}\}$-plane, overlaid on the equilibrium curves. The plots show how the system goes through the irreversible sequence of states. These state transitions are made possible by exploiting the following critical effect. When the state trajectory follows a stable branch and reaches a point where it becomes unstable, the trajectory rapidly "jumps" and converges to a stable branch, moving very close to a cubic arc of constant total volume. Figure \ref{fig:TwoChamber} shows the flow velocity and pressure distribution corresponding to an attractive instance in which the system passes through point A, presented in time and on the $\{\lambda_{2},\lambda_{1}\}$-plane, in Figure \ref{fig:ODE45}. When the system reaches point A, the flow rate from the first chamber (the one connected to the inlet tube) towards the second chamber is spontaneously bigger than the inlet flow rate. Therefore, the first chamber is deflated.
\section{Concluding Remarks}
\label{Concludings}
In this work, we analyzed the dynamics of creeping flow in a bi-stable hyperelastic spherical chamber by calculating the velocity field and pressure distribution inside the chamber. The analytical results were compared to numerical simulations, which showed an excellent fit. From the normalization of the governing equations (\ref{momentum_tube}) and (\ref{Hagen-Poiseuille}), we obtained the condition $\varepsilon=a^4/r_{0}^3\ell\ll1$ which led to creeping flow inside the chambers. Moreover, we obtained the condition $q^{*}\ll \mu a/\rho$ for inertial effects to be negligible in the vicinity of the connection to the tubes.

In order to describe the coupled model of viscous-elastic dynamics, we first focused on the non-linear constitutive elastic laws (Mooney-Rivlin model). Next, using the analytical series solution, we studied the dynamic responses for two physical different cases. The first case was dictated volumetric flux. Based on the mechanical energy principle, we formulated the chamber's pressure distribution. We obtained the characteristic pressure in the elastic chambers as $p^{*}=w_{0}\psi^{*}/t_{0}$ which depends on the hyperelastic model we used.

The pressure is spatially uniform at the leading order. Mainly, the leading order of the problem is a case of fully developed uniform pressure without any velocities. More spatial and temporal pressure and velocities are created in the dynamic case where the chamber is inflated or deflated. Based on the numeric simulation, we saw that close to the chamber's wall ($\chi\rightarrow 1$), the viscous flow will generate $O(\epsilon_{t})$ spatially varying corrections.
In this analysis, we obtained a non-intuitive result - the pressure solution on the chamber's wall obtained its maximum value at an intermediate value of $\theta$ away from the poles $\theta=0$ and $\theta=\pi$. This qualification may be critical in identifying the failure point of the chamber's wall. 

In the second physical case, the flow was driven by an imposed inlet pressure. There we have simplified and reduced the time-varying dynamical equations to one compact equation depending on  the elastic model, dictated pressure, chamber stretch, and inlet tube slenderness (\ref{ODEsimplified}). Although the finite element simulation showed that the chamber undergoes a slightly different deformation from an ideal sphere (making it slightly pear-shaped), the effect is small and the solution obtained in this work are excellent approximation. However, in order to characterize the obtained geometric shape, more extensive analysis is required, which goes beyond the scope of this work.

In the last part of this work, we presented an investigation of a system consisting of two interconnected coupled chambers controlled by the flow at the chamber's inlet. This system demonstrates the bi-stability feature through which the chambers can display controlled transitions between different multi-stable states using a single input of controlled flow rate.

We have developed an analytical model that can serve as the basis for the physical understanding of acinar fluid mechanics in a rhythmically expanding spherical alveolus and its vicinity. These results allow modeling of flows in applications such as the inflation of balloons in medical procedures and pulmonary drug delivery optimization to target specific lung regions. Moreover, the results might be leveraged to analyze the dynamics of particles inside spherical elastic chambers in low-Reynolds flow as future work. 
\newpage
\appendix
\section{Asymptotic approximations for the equilibrium equation $P_{SS}(\lambda_{SS})$}\label{appA}
In section \ref{ElsticSection} the relation between stretch, $\lambda_{SS}$, and pressure, $P_{SS}$, in equilibrium condition has been presented. This well-known relation was extensively leveraged to describe the quasi-static inflation of spherical balloons \citep{Beatty,Treloar,BenHaim} for spatially uniform pressures. As seen from Figure \ref{fig:staticCase}(a), showing the relation in (\ref{Pss}) with $\alpha=0.1$, the uniform pressure of the chamber, $P_{SS}(\lambda_{SS})$ has two bifurcation points, described by a local maximum point at $(\lambda_{A},P_{A})$, and a local minimum point at $(\lambda_{B},P_{B})$. This figure shows a bifurcation, which occurs when the pressure enters or exits the range between the local extrema, $P_{A}<P_{SS}<P_{B}$, illustrated in grey. Here, we shall obtain an asymptotic approximations for the bifurcation points of the equilibrium curve, $P_{A},P_{B},\lambda_{A}$ and $\lambda_{B}$.
Since the static behavior of the system is dependent on the value of the uniform pressure, the bifurcation points $(\lambda_{A},P_{A})$ and $(\lambda_{B},P_{B})$ may be computed. These extrema are the roots of the derivative of the system energy, given by the roots of the quartic polynomial equation,
\begin{equation}
\label{equaX}
    x^{3}-7=\alpha(x^{4}+5x),
\end{equation}
where $x=\lambda_{SS}^{2}$. Since $\alpha\ll1$ we use the iterative asymptotics, yielding,   
\begin{equation}
    x_{i+1}=\sqrt[3]{7+\alpha(x_{i}^{4}+5x_{i}}).
\end{equation}
Starting with $x_{0}=\sqrt[3]{7}$ which is obtained from the leading-order solution ($\alpha=0$), after two iterative steps, the approximation is obtained as,
\begin{equation}
\label{lambdaA}
    \lambda_{A}=\sqrt{x_{2}}=\sqrt[6]{7}+\cfrac{2}{\sqrt{7}}\alpha+\frac{12}{7\sqrt[6]{7}}\alpha^2+O(\alpha^3).
\end{equation}
Substitution of (\ref{lambdaA}) into (\ref{Pss}) and then using the Taylor series approximation, yields the regular approximation for the local maximum point as,
\begin{equation}
\label{PAapprox}
    P_{A}=\cfrac{24}{7\sqrt[6]{7}}+\frac{24}{\sqrt[6]{7^5}}\alpha+\frac{48}{7\sqrt{7}}\alpha^2+O(\alpha^3).
\end{equation}\\
In order to approximate the local minimum point, we  use another singular asymptotic method. We consider $y=\alpha x$ and rewrite the equation (\ref{equaX}) as
\begin{equation}
\label{equaY}
    y^{4}-y^{3}+\alpha^{3}(5y+7)=0.
\end{equation}
Since the new equation is regular, we approximate the minimum point by asymptotic expansion with regard to the small parameter $\alpha$, 
\begin{equation}
\label{PBappr}
    y(\alpha)=1+\alpha y_{1}+\alpha^{2} y_{2}+\alpha^{3}y_{3}+O(\alpha^{4}).
\end{equation}\\
This expansion is formally substituted into the algebraic equation (\ref{equaY}), and the coefficients of the powers of $\alpha$ are compared. Then, the approximation of $\lambda_{B}$ and $P_{B}$ are obtained,
\begin{equation}
\label{PBappr}
    \begin{split}
        \lambda_{B}&=\frac{1}{\sqrt{\alpha}}+6\alpha^{3}\sqrt{\alpha}+O(\alpha^{6}),\\[5pt]
        P_{B}&=8\sqrt{\alpha}-8\sqrt{\alpha^{7}}+O(\sqrt{\alpha^{15}}).
    \end{split}
\end{equation}
The  evolution  of  those  extrema as a function of the small parameter $\alpha$ is presented in Figure \ref{fig:staticCase}(d). This figure shows a further bifurcation, implying that a region with three equilibria exist only when $0<\alpha<0.214$. Those parameters are considered to lie within the physical domain of rubber-like materials, as well as for biological tissues.\\
The next curve we examine is the solution of the equilibrium equation (\ref{Pss}). Since the equation has no analytical solution, we shall find an asymptotic approximation by separating the solution into three main regions, as shown in Figure \ref{fig:staticCase}(c). In the first region, $(I)$, the pressure in the chamber is lower than the maximum point, $0<P_{SS}<P_{A}$ and $1<\lambda_{SS}<\lambda_{A}$. In the second region, $(II)$, the pressure in the chamber is between the minimum and maximum points of the graph, $P_{B}<P_{SS}<P_{A}$ and $\lambda_{A}<\lambda_{SS}<\lambda_{B}$. In the third region, $(III)$, the pressure in the chamber is higher than the minimum point on the graph, $P_{SS}>P_{B}$ and $\lambda_{B}<\lambda_{SS}$.
In the first case (where $P<P_{A}$), the equilibrium point will be close to $\lambda_{SS}=1$. \cite{Ilssar} present the following approximation to describe this branch by $\lambda_{SS}=1+\delta_{1}+\delta_{1}^{2}+O(\delta_{1}^{3})$, where $\delta_{1}$ is formulated by,
\begin{equation}
\label{delta1Appr}
    \delta_{1}(P_{SS};\alpha)=\cfrac{-7P_{SS}+24+24\alpha-\sqrt{3\big[64\alpha P_{SS}+192(\alpha+1)^{2}-21P_{SS}^{2}\big]}}{8(7P_{SS}-33\alpha-21)}.
\end{equation}
Next, the second stable equilibrium radius, which exists when the pressure is higher than the local minimum ($P>P_{B}$), is considered significantly larger than unity. Thus, in order to formulate an approximation for this equilibrium stretch, $\lambda_{SS}=\delta_{2}^{-1}(\alpha)\gg 1$ is substituted into (\ref{Pss}). After some regular algebraic manipulation, a second-order algebraic equation in terms of $\lambda_{SS}$ is provided. The solution is given by
\begin{equation}
\label{delta2Appr}
    \delta_{2}(P_{SS};\alpha)=\cfrac{P_{SS}\pm\sqrt{P_{SS}^{2}-64\alpha}}{8}.
\end{equation}
The solution having a positive sign in front of the square root of the discriminant, is suitable for the solution of the third case in which $P_{SS}>P_{B}$, while the solution with the negative sign is suitable for solution of the unstable branch in which $P_{B}<P_{SS}<P_{A}$. Those approximations are also plotted in Figure \ref{fig:staticCase}(c) with dash-lines on the solid curve representing the exact solution.

\section{Asymptotic justification for neglecting the non-spherical deformation of the chamber}\label{appB}
For the case of a narrow tube filling a larger chamber, the pressure within the chamber involves a large spatially uniform part and a small order correction, $\varepsilon P_{1}(R,\theta;\tilde{\epsilon};T)$. This result was obtained analytically by (\ref{GeneralSolution}) without the assumption of spherical deformation. Assuming that the chamber's radius is large relative to the tube's radius ($\epsilon_{a}\ll 1$), and based on the coupled model of fluid-solid numerical simulation results (see Figure. \ref{fig:SingleInputVelocitiesCFD}), it is clear that the maximum pressure gradient is obtained in the inlet (or outlet) region and significantly decayed within the chamber's area. Since far away from the tube ($\chi\rightarrow 1$), the pressure is approximately hydrostatic, the real shape of the chamber will consist of an ideal sphere in addition to a small perturbation. Based on the numerical results we have obtained in § \ref{Dynamic_case_II}, the perturbation of the chamber's stretch is $O(\epsilon_{t})$, so we shall assume a regular asymptotic approximation as follow,
\begin{equation}
\label{delta2Appr}
      \lambda(\theta;T)=\lambda_{0}(T)+\epsilon_{t}\lambda_{1}(\theta;T)+O(\epsilon_{t}^2).
\end{equation}
As we have already seen in section § \ref{sec_Solution}, the pressure is spatially uniform at the leading order. Mainly, the leading order of the problem is a case of fully developed uniform pressure without any velocities. According to the solution obtained by \cite{Beatty}, $P_{S}(\lambda_{0})=\lambda_{0}^{-2}\cdot d\psi/d\lambda_{0}$ is the well-known isotropic pressure which represents the hydrostatic pressure in the leading order. On the other hand, from the CFD simulation, we saw that close to the chamber's wall ($\chi\rightarrow 1$), the viscous flow will generate $O(\epsilon_{t})$ spatially varying corrections. Therefore, it is convenient to assume that pressure distribution can also be approximated as
\begin{equation}
\label{MoreSeries}
    P(R,\theta;T)\approx P_{S}(\lambda_{0})+\epsilon_{t}\tilde{P}(R,\theta;T)\quad \text{at} \quad \chi\rightarrow 1,
\end{equation}
where $\tilde{P}\thicksim O(1)$ and $\epsilon_{t}\tilde{P}\thicksim\varepsilon P_{1}$ close to the inner chamber's wall. In order to examine the effect of non-spherical deformation on the resulting pressure profile (without considering nonlinear hyperelastic analysis that is outside the scope of this work), we wish to use the regular asymptotic approximation (\ref{MoreSeries}) and set the radial perturbation (\ref{delta2Appr}),
\begin{equation}
\label{integralp}
   P(R=\lambda(\theta;T),\theta;T)\approx P_{S}(\lambda_{0})+\epsilon_{t} \tilde{P}(R=\lambda(\theta;T),\theta;T)\quad \text{at} \quad \chi\rightarrow 1,
\end{equation}
by taking the Taylor approximate, we get
\begin{equation}
\label{integralp}
   P(R=\lambda(\theta;T),\theta;T)=P_{S}(\lambda_{0})+\epsilon_{t}\tilde{P} +O(\epsilon_{t}^2) \thicksim P_{S}+\varepsilon P_{1};
\end{equation}
hence, the $O(\epsilon_{t})$ in the chamber's shape create $O(\epsilon_{t}^2)$ pressure correction.
In order to find the full fluid-structure interaction, the entire elastic equations must be solved coupled with the Stokes equation, but this analysis will be performed as future work. 
\newpage
\bibliographystyle{jfm}
\bibliography{jfm-instructions}

\begin{thebibliography}{26}
\expandafter\ifx\csname natexlab\endcsname\relax\def\natexlab#1{#1}\fi
\def\au#1{#1} \def\ed#1{#1} \def\yr#1{#1}\def\at#1{#1}\def\jt#1{\textit{#1}}
  \def\bt#1{#1}\def\bvol#1{\textbf{#1}} \def\vol#1{#1} \def\pg#1{#1}
  \def\publ#1{#1}\def\arxiv#1{#1}\def\org#1{#1}\def\st#1{\textit{#1}}

\bibitem[Beatty(1987)]{Beatty}
{\sc \au{Beatty, Millard~F}} \yr{1987}  \at{Topics in finite elasticity:
  hyperelasticity of rubber, elastomers, and biological tissues—with
  examples} .

\bibitem[Ben-Haim {\em et~al.\/}(2020)Ben-Haim, Salem, Or \& Gat]{BenHaim}
{\sc \au{Ben-Haim, E.}, \au{Salem, L.}, \au{Or, Y.} \& \au{Gat, A.~D.}}
  \yr{2020}  \at{Single-input control of multiple fluid-driven elastic
  actuators via interaction between bistability and viscosity}.  \jt{Soft
  Robot}  \bvol{7}~(2),  \pg{259--265}.

\bibitem[Davidson \& Fitz-Gerald(1972)]{Davidson}
{\sc \au{Davidson, MR} \& \au{Fitz-Gerald, JM}} \yr{1972}  \at{Flow patterns in
  models of small airway units of the lung}.  \jt{Journal of Fluid mechanics}
  \bvol{52}~(1),  \pg{161--177}.

\bibitem[Dreyer {\em et~al.\/}(1982)Dreyer, Müller \& Strehlow]{Dreyer}
{\sc \au{Dreyer, W}, \au{Müller, I} \& \au{Strehlow, P}} \yr{1982}  \at{A
  study of equilibria of interconnected balloons}.  \jt{The Quarterly Journal
  of Mechanics and Applied Mathematics}  \bvol{35}~(3),  \pg{419--440}.

\bibitem[Elbaz \& Gat(2014)]{Elbaz14}
{\sc \au{Elbaz, S.~B.} \& \au{Gat, A.~D.}} \yr{2014}  \at{Dynamics of viscous
  liquid within a closed elastic cylinder subject to external forces with
  application to soft robotics}.  \jt{Journal of Fluid Mechanics}  \bvol{758},
  \pg{221--237}.

\bibitem[Elbaz \& Gat(2016)]{Elbaz16}
{\sc \au{Elbaz, S.~B.} \& \au{Gat, A.~D.}} \yr{2016}  \at{Axial creeping flow
  in the gap between a rigid cylinder and a concentric elastic tube}.
  \jt{Journal of Fluid Mechanics}  \bvol{806},  \pg{580--602}.

\bibitem[Fei \& Gao(2014)]{Fei14}
{\sc \au{Fei, Yanqiong} \& \au{Gao, Hanwei}} \yr{2014}  \at{Nonlinear dynamic
  modeling on multi-spherical modular soft robots}.  \jt{Nonlinear Dynamics}
  \bvol{78}~(2),  \pg{831--838}.

\bibitem[Fei \& Pang(2016)]{Fei16}
{\sc \au{Fei, Yanqiong} \& \au{Pang, Wu}} \yr{2016}  \at{Analysis on nonlinear
  turning motion of multi-spherical soft robots}.  \jt{Nonlinear Dynamics}
  \bvol{88}~(2),  \pg{883--892}.

\bibitem[Gamus {\em et~al.\/}(2017)Gamus, Salem, Ben-Haim, Gat \& Or]{Gamus}
{\sc \au{Gamus, Benny}, \au{Salem, Lior}, \au{Ben-Haim, Eran}, \au{Gat, Amir~D}
  \& \au{Or, Yizhar}} \yr{2017}  \at{Interaction between inertia, viscosity,
  and elasticity in soft robotic actuator with fluidic network}.  \jt{IEEE
  Transactions on Robotics}  \bvol{34}~(1),  \pg{81--90}.

\bibitem[Glozman {\em et~al.\/}(2010)Glozman, Hassidov, Senesh \&
  Shoham]{Glozman}
{\sc \au{Glozman, Daniel}, \au{Hassidov, Noam}, \au{Senesh, Merav} \&
  \au{Shoham, Moshe}} \yr{2010}  \at{A self-propelled inflatable earthworm-like
  endoscope actuated by single supply line}.  \jt{IEEE Transactions on
  Biomedical Engineering}  \bvol{57}~(6),  \pg{1264--1272}.

\bibitem[Gorissen {\em et~al.\/}(2019)Gorissen, Milana, Baeyens, Broeders,
  Christiaens, Collin, Reynaerts \& De~Volder]{Gorissen}
{\sc \au{Gorissen, B.}, \au{Milana, E.}, \au{Baeyens, A.}, \au{Broeders, E.},
  \au{Christiaens, J.}, \au{Collin, K.}, \au{Reynaerts, D.} \& \au{De~Volder,
  M.}} \yr{2019}  \at{Hardware sequencing of inflatable nonlinear actuators for
  autonomous soft robots}.  \jt{Adv Mater}  \bvol{31}~(3),  \pg{e1804598}.

\bibitem[Haber {\em et~al.\/}(2000)Haber, Butler, Brenner, Emanuel \&
  Tsuda]{Haber}
{\sc \au{Haber, S.}, \au{Butler, J.~P.}, \au{Brenner, H.}, \au{Emanuel, I.} \&
  \au{Tsuda, A.}} \yr{2000}  \at{Shear flow over a self-similar expanding
  pulmonary alveolus during rhythmical breathing}.  \jt{Journal of Fluid
  Mechanics}  \bvol{405},  \pg{243--268}.

\bibitem[Happel \& Brenner(2012)]{Happel}
{\sc \au{Happel, John} \& \au{Brenner, Howard}} \yr{2012} {\em Low Reynolds
  number hydrodynamics: with special applications to particulate media\/}, ,
  \vol{vol.~1}.  \publ{Springer Science and Business Media}.

\bibitem[Hines {\em et~al.\/}(2017)Hines, Petersen \& Sitti]{Hines}
{\sc \au{Hines, Lindsey}, \au{Petersen, Kirstin} \& \au{Sitti, Metin}}
  \yr{2017}  \at{Asymmetric stable deformations in inflated dielectric
  elastomer actuators}  \pg{pp. 4326--4331}.

\bibitem[Ilssar \& Gat(2020)]{Ilssar}
{\sc \au{Ilssar, Dotan} \& \au{Gat, Amir~D}} \yr{2020}  \at{On the inflation
  and deflation dynamics of liquid-filled, hyperelastic balloons}.  \jt{Journal
  of Fluids and Structures}  \bvol{94},  \pg{102936}.

\bibitem[Manfredi {\em et~al.\/}(2019)Manfredi, Capoccia, Ciuti \&
  Cuschieri]{Manfredi}
{\sc \au{Manfredi, Luigi}, \au{Capoccia, Elisabetta}, \au{Ciuti, Gastone} \&
  \au{Cuschieri, Alfred}} \yr{2019}  \at{As oft p neumatic i nchworm d ouble
  balloon (spid) for colonoscopy}.  \jt{Scientific reports}  \bvol{9}~(1),
  \pg{1--9}.

\bibitem[Mangan \& Destrade(2015)]{Mangan}
{\sc \au{Mangan, Robert} \& \au{Destrade, Michel}} \yr{2015}  \at{Gent models
  for the inflation of spherical balloons}.  \jt{International Journal of
  non-linear mechanics}  \bvol{68},  \pg{52--58}.

\bibitem[Milic-Emili {\em et~al.\/}(1964)Milic-Emili, Mead, Turner \&
  Glauser]{Milic-Emili}
{\sc \au{Milic-Emili, JJJE}, \au{Mead, JTURNERJM}, \au{Turner, JM} \&
  \au{Glauser, EM}} \yr{1964}  \at{Improved technique for estimating pleural
  pressure from esophageal balloons}.  \jt{Journal of Applied Physiology}
  \bvol{19}~(2),  \pg{207--211}.

\bibitem[Needleman(1977)]{Needleman}
{\sc \au{Needleman, Alan}} \yr{1977}  \at{Inflation of spherical rubber
  balloons}.  \jt{International Journal of Solids and Structures}
  \bvol{13}~(5),  \pg{409--421}.

\bibitem[Ogden(1972)]{Ogden}
{\sc \au{Ogden, Raymond~William}} \yr{1972}  \at{Large deformation isotropic
  elasticity–on the correlation of theory and experiment for incompressible
  rubberlike solids}.  \jt{Proceedings of the Royal Society of London. A.
  Mathematical and Physical Sciences}  \bvol{326}~(1567),  \pg{565--584}.

\bibitem[Overvelde {\em et~al.\/}(2015)Overvelde, Kloek, D'Haen~J \&
  Bertoldi]{Overvelde}
{\sc \au{Overvelde, J.~T.}, \au{Kloek, T.}, \au{D'Haen~J, J.} \& \au{Bertoldi,
  K.}} \yr{2015}  \at{Amplifying the response of soft actuators by harnessing
  snap-through instabilities}.  \jt{Proc Natl Acad Sci U S A}  \bvol{112}~(35),
   \pg{10863--8}.

\bibitem[Salem {\em et~al.\/}(2020)Salem, Gamus, Or \& Gat]{Salem}
{\sc \au{Salem, Lior}, \au{Gamus, Benny}, \au{Or, Yizhar} \& \au{Gat, Amir~D}}
  \yr{2020}  \at{Leveraging viscous peeling to create and activate soft
  actuators and microfluidic devices}.  \jt{Soft Robotics}  \bvol{7}~(1),
  \pg{76--84}.

\bibitem[Siefert {\em et~al.\/}(2019)Siefert, Reyssat, Bico \& Roman]{Siefert}
{\sc \au{Siefert, Emmanuel}, \au{Reyssat, Etienne}, \au{Bico, José} \&
  \au{Roman, Benoit}} \yr{2019}  \at{Bio-inspired pneumatic shape-morphing
  elastomers}.  \jt{Nature materials}  \bvol{18}~(1),  \pg{24--28}.

\bibitem[Treloar(1975)]{Treloar}
{\sc \au{Treloar, Leslie Ronald~George}} \yr{1975} {\em The physics of rubber
  elasticity\/}.  \publ{Oxford University Press, USA}.

\bibitem[Vandermarlière(2016)]{Vandermarliere}
{\sc \au{Vandermarlière, Julien}} \yr{2016}  \at{On the inflation of a rubber
  balloon}.  \jt{The Physics Teacher}  \bvol{54}~(9),  \pg{566--567}.

\bibitem[Yamamoto {\em et~al.\/}(2001)Yamamoto, Sekine, Sato, Higashizawa,
  Miyata, Iino, Ido \& Sugano]{Yamamoto}
{\sc \au{Yamamoto, H.}, \au{Sekine, Y.}, \au{Sato, Y.}, \au{Higashizawa, T.},
  \au{Miyata, T.}, \au{Iino, S.}, \au{Ido, K.} \& \au{Sugano, K.}} \yr{2001}
  \at{Total enteroscopy with a nonsurgical steerable double-balloon method}.
  \jt{Gastrointest Endosc}  \bvol{53}~(2),  \pg{216--20}.

\end{thebibliography}
\end{document}